\documentclass[11pt]{article} 
\usepackage[utf8]{inputenc}

\usepackage{amsfonts}
\usepackage{amsmath}
\usepackage{amssymb}
\usepackage{amsthm}
\usepackage{caption}
\usepackage{color}
    \definecolor{darkgreen}{rgb}{0,0.5,0}
    \definecolor{darkblue}{rgb}{0,0,0.6}
    \definecolor{purple}{rgb}{0.4,.2,0.7}
\usepackage[margin = 2.5cm]{geometry}
\usepackage{graphicx}
\usepackage[hyperfootnotes = false, colorlinks = true, linkcolor = darkblue, citecolor = purple]{hyperref}
\usepackage{subcaption}
\usepackage{xfakebold}
    \newcommand{\fbseries}{\unskip\setBold\aftergroup\unsetBold\aftergroup\ignorespaces}
    \makeatletter
    \newcommand{\setBoldness}[1]{\def\fake@bold{#1}}
    \makeatother
    \setBoldness{0.6}

\newcommand{\be}{\begin{equation}}
\newcommand{\ee}{\end{equation}}
\def\la{\label}
\def\nref#1{(\ref{#1})}

\begin{document}

\thispagestyle{empty}
\begin{center}
    ~\vspace{5mm}
    
    {\LARGE \bf {Islands outside the horizon \\}} 
    
    \vspace{0.5in}
    
    {\bf Ahmed Almheiri,$^1$ Raghu Mahajan,$^{1,2}$ Juan Maldacena$^1$}

    \vspace{0.5in}

    $^1$Institute for Advanced Study,  Princeton, NJ 08540, USA \vskip1em
    $^2$Jadwin Hall, Princeton University, Princeton, NJ 08540, USA
    
    \vspace{0.5in}
    
    {\tt almheiri@ias.edu, raghu\_m@princeton.edu, malda@ias.edu}
\end{center}

\vspace{0.5in}

\begin{abstract}
We consider an AdS$_2$ black hole in equilibrium with a bath, which we take to have a dual description as (0+1)-dimensional quantum mechanical system coupled to a (1+1)-dimensional field theory serving as the bath. 
We compute the entropies of both the quantum mechanical degrees of freedom and of the bath separately, while allowing contributions from entanglement wedge ``islands''. 
We find situations where the island extends {\it outside} the black hole horizon. 
This suggests possible causality paradoxes which we show are avoided because of the quantum focusing conjecture.
Finally, we formulate a version of the information paradox for a black hole in contact with a bath in the Hartle-Hawking state, and demonstrate the role of islands in resolving this paradox.
\end{abstract}

\vspace{1in}

\pagebreak

\setcounter{tocdepth}{3}
{\hypersetup{linkcolor=black}\tableofcontents}

\section{Introduction} 

The emergence of a Page curve for evaporating black holes is associated to contributions to the von Neumann entropy of Hawking radiation coming from an \emph{island} behind the horizon \cite{Penington:2019npb,Almheiri:2019psf,Almheiri:2019hni}. 
The prescription is to compute the entropy of the union of the Hawking radiation and the island, plus a contribution from the boundary area of the island:
\begin{align}
S[\mathrm{ \fbseries Rad}] ={\rm min} \left\{ {\rm ext} \left[  S[\mathrm{Rad}\cup I] + {\mathrm{Area}[\partial I] \over 4 G_N} \right]\right\}\, .
\label{islandrule}
\end{align}
Here, by ``{\fbseries Rad}'' we mean the radiation in the full quantum description. 
In contrast, by ``Rad'' we mean the description of radiation in the semiclassical description.  
We extremize the generalized entropy over all possible islands $I$, and the minimum value gives the entropy of Hawking radiation.
The reason that (\ref{islandrule}) leads to a decreasing fine-grained entropy for the Hawking radiation is that the system $\mathrm{Rad} \cup I$ contains both the outgoing Hawking radiation and its interior purifying partner.

These islands could in principle exist anywhere in the spacetime, with potentially interesting consequences if they exist inside the causal wedge. 
In previously considered examples \cite{Penington:2019npb, Almheiri:2019psf,  
Almheiri:2019hni}, the islands were always behind a black hole horizon.
In this work, we will discuss examples where the islands exist {\it outside} black hole horizons. 
These new examples involve black holes that are eternally coupled to, and in equilibrium with, a bath at zero or non-zero temperatures.  

Following \cite{Almheiri:2019psf}, we will consider nearly-AdS$_2$ black holes in Jackiw-Teitelboim gravity with conformal matter, which are eternally joined along the AdS boundary to an external flat-space CFT.
This can be thought of as a toy model for asymptotically-flat near-extremal Reissner-Nordstr\"om black holes, where we neglect the effects of gravity in the asymptotically-flat region. 
The goal is to analyze the role of islands in the computation of the entropy of various regions. 
An interesting result we find is that the near-horizon region outside the black hole is contained in an island corresponding to a region far away in the bath. 
This suggests that the physics of the near-horizon region is encoded in the state of the bath far away from the black hole.

This observation leads to the counterintuitive result that the entanglement wedge associated to the dual quantum system at the AdS$_2$ boundary is contained within  its causal wedge.
This gives rise to an apparent paradox: If we were to decouple the black hole from the bath, could  the boundary system  get a signal from the island? 
The naive spacetime diagram suggests that we do.
However, we expect that the decoupling process will produce some nonzero energy which should move the horizon outwards so that the island lies behind the final horizon.
We argue that the quantum focusing conjecture (QFC) \cite{Bousso:2015mna} is sufficient to avoid a paradox.
 
We then consider a black hole in equilibrium with a bath at finite temperature. 
The system can be purified by introducing the thermofield double, which is described by the Hartle-Hawking state in the two-sided geometry \cite{Israel:1976ur}.
In this system, it is now possible to formulate a version of the Hawking information paradox \cite{Hawking:1976ra}, or its AMPS version \cite{Almheiri:2012rt}, by looking at the behavior of the entropy of the pair of boundary quantum mechanical systems or the pair of outside regions.  
A very closely related paradox was discussed in \cite{Mathur:2014dia}. 
The system is invariant under boosts, which involve forwards time translations on one side and backwards on the other.
However, it has interesting time dependence if we consider forwards time evolution on both sides. 
We will analyze the evolution of the entanglement entropy of the union of the two boundary quantum mechanical systems.
Initially, the entropy of the union of the two boundaries is relatively small because the two boundaries are mostly  entangled with each other. 
As we evolve forwards on both sides, the black holes emit Hawking radiation and absorb radiation from the bath. 
There is no gravitational backreaction because the energy that is emitted precisely balances the energy that falls in.  
However, this process leads to a growing entanglement between each boundary system and the bath, which causes the entropy of the pair of boundaries to grow linearly with time. 
But the entropy of the pair of the boundaries should not be larger than twice their Bekenstein-Hawking entropy. 
This apparent paradox is resolved by the appearance of islands. 
At late times, we get an island that spans the black hole interior and also invades a little outside each of the two horizons. 
Once the island is present, the entropy of the boundaries is constant in time and is given essentially by twice the Bekenstein-Hawking answer.
The final resolution of this version of the information paradox is conceptually as in \cite{Penington:2019npb,Almheiri:2019psf}, but it is technically simpler because the geometry is that of the usual eternal black hole, and we do not have to find the time-dependent geometry of an evaporating black hole.

This paper is organized as follows. 
In section \ref{sec-zeroT}, we discuss the basic setup of a two-dimensional gravity theory coupled to a bath. 
We consider the zero temperature case and compute the entanglement wedge of a region including the boundary quantum mechanical system. 
We find a quantum extremal surface outside the horizon. 
We also discuss how this is consistent with causality, thanks to the quantum focusing conjecture. 
In section \ref{sec-finiteT}, we consider the same system but at finite temperature. Again we get a quantum extremal surface outside the horizon. 
In section \ref{sec:eternalinfo}, we discuss a version of the information paradox for the Hartle-Hawking state of a black hole coupled to a bath.
In section \ref{sec:pheno}, we discuss applications of our ideas to some four-dimensional black holes, and also to the transplanckian problem in Hawking radiation and cosmology.
We conclude in section \ref{sec-discussion} with some discussion.

\section{Extremal black hole coupled to a bath} 
\label{sec-zeroT}

In this section, we consider an extremal black hole coupled to a bath. 
By a bath we mean a quantum mechanical system with no gravity, or one where we can neglect the effects of gravity.
This setup could be an approximation to a black hole in asymptotically-flat space, where we neglect the effects of quantum gravity in the flat space region far away.

\begin{figure}[t]
    \begin{center}
    \includegraphics[scale=.4]{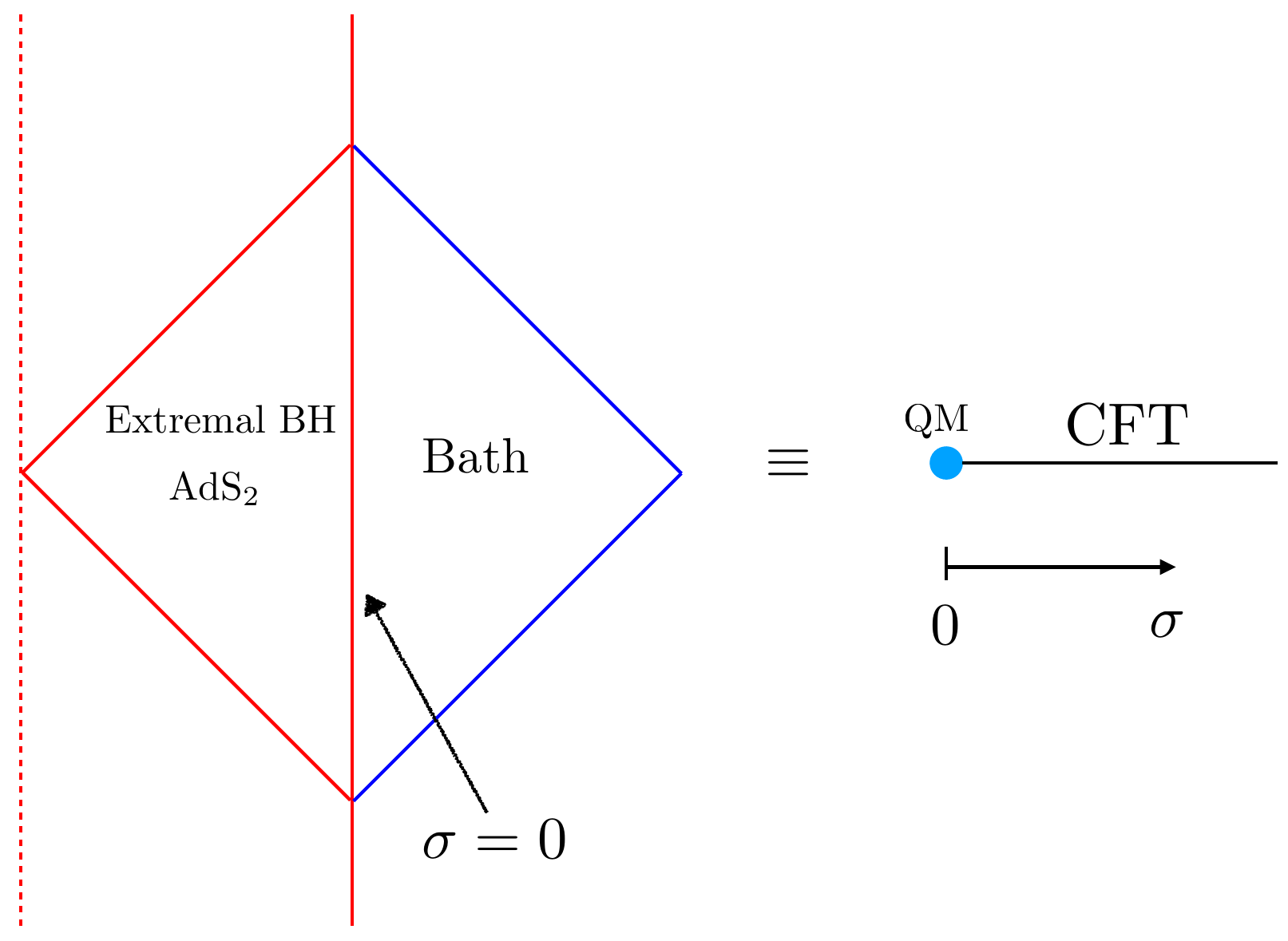}
    \end{center}
    \caption{On the left, we depict the Penrose diagram for a two-dimensional extremal black
    hole, coupled to a bath at zero temperature. 
    The red part covers a region of AdS$_2$.
    The blue triangle is the Penrose diagram of half of Minkowski space. 
    On the right, we have the dual quantum mechanical setup, with the black hole replaced by a quantum system that is coupled to a CFT$_2$ on a half-line (only the spatial dimension is represented).}
    \label{ZeroTSetup}
\end{figure}
More specifically, we consider two-dimensional Jackiw-Teitelboim gravity, with matter given by a two-dimensional CFT with central charge $c$.
Consider a zero-temperature black hole in this theory. 
We now couple this gravity theory to a bath that consists of the same CFT$_2$, but living in half of Minkowski space.
We couple them by imposing ``transparent'' boundary conditions at the locus where the boundary of AdS$_2$ joins the half-line where the bath lives, see figure \ref{ZeroTSetup}.
The full action is 
\be
  I =  { 1 \over 4 \pi } \int d^2x\,  \sqrt{-g}\, [\phi R + 2(\phi-\phi_0)] + 
I_\text{CFT}\, .
\ee
Here, we have set $4G_N=1$, and $\phi_0$ is a constant that gives rise to the extremal entropy. 
We impose the usual JT boundary conditions on the metric $g_{uu} = \frac{1}{\varepsilon^2}$ and dilaton $\phi - \phi_0 = \frac{\phi_r}{\varepsilon}$.
The gravity system is only on half the space, but the CFT is over the whole line.
Thus the CFT is partly in the dynamical gravity region, and partly in the half space with a fixed flat metric.

The full system can be described by the Penrose diagram shown in figure \ref{ZeroTSetup}.
We denote by $y^\pm = t \pm \sigma$ the coordinates in the bath region, with $\sigma >0$. 
The black hole can be described using the Poincar\'e coordinates $x^\pm = x^0 \pm x^1 $, with $x^1<0$.
The metric and the dilaton profiles in the gravity region are
\be \la{AdSTm}
d s^2 = {- 4  \, d x^+ dx^- \over (x^- -x^+)^2 } \, , \quad \phi =\phi_0 +  { 2 \phi_r \over (x^- -x^+ ) } \, .
\ee

The bulk state is taken to be in the Poincar\'e vacuum $T^{(x)}_{++} = T^{(x)}_{--} = 0$. 
Since the black hole and the bath are taken to be in equilibrium, an extremal black hole requires a bath with zero temperature.
Therefore we must have a vanishing bath stress tensor $T^{(y)}_{++} = T^{(y)}_{--} = 0$. 
The transparent boundary condition implies that the bulk and bath stress tensors are related as
\begin{align}
\left( {\partial x^+ \over \partial y^+ }\right)^2T^{(x)}_{++} =  T^{(y)}_{++} + {c \over 24 \pi} \{ x^+, y^+\} \, ,
\la{StressT}
\end{align}
and an identical equation for the right-moving piece.
Since both $T^{(x)}_{\mu\nu}$ and $T^{(y)}_{\mu\nu}$ vanish, we require that $x^+(y^+)$ and $x^-(y^-)$ be $SL(2,\mathbb{R})$ transformations.
The requirement of having a pure state of the bath plus bulk fixes this transformation, up to an overall scale, to be simply $x^+ = y^+$ and $x^- = y^-$. 
The stress tensor of matter is zero everywhere. 

The gravitational part of the system can have a dual quantum mechanical description. 
We view this dual quantum system as living at $\sigma=0$, and coupled to the CFT that lives in the half-space $\sigma > 0$, see figure \ref{ZeroTSetup}. 
We will not need the explicit form of this quantum mechanical system. 
When we talk about the full microscopic von Neumann entropies of the quantum state, we imagine that they are defined in this combined quantum system, the one living at $\sigma=0$ and the bath CFT living on $\sigma>0$.

\begin{figure}[t!]
    \begin{center}
    \includegraphics[scale=.4]{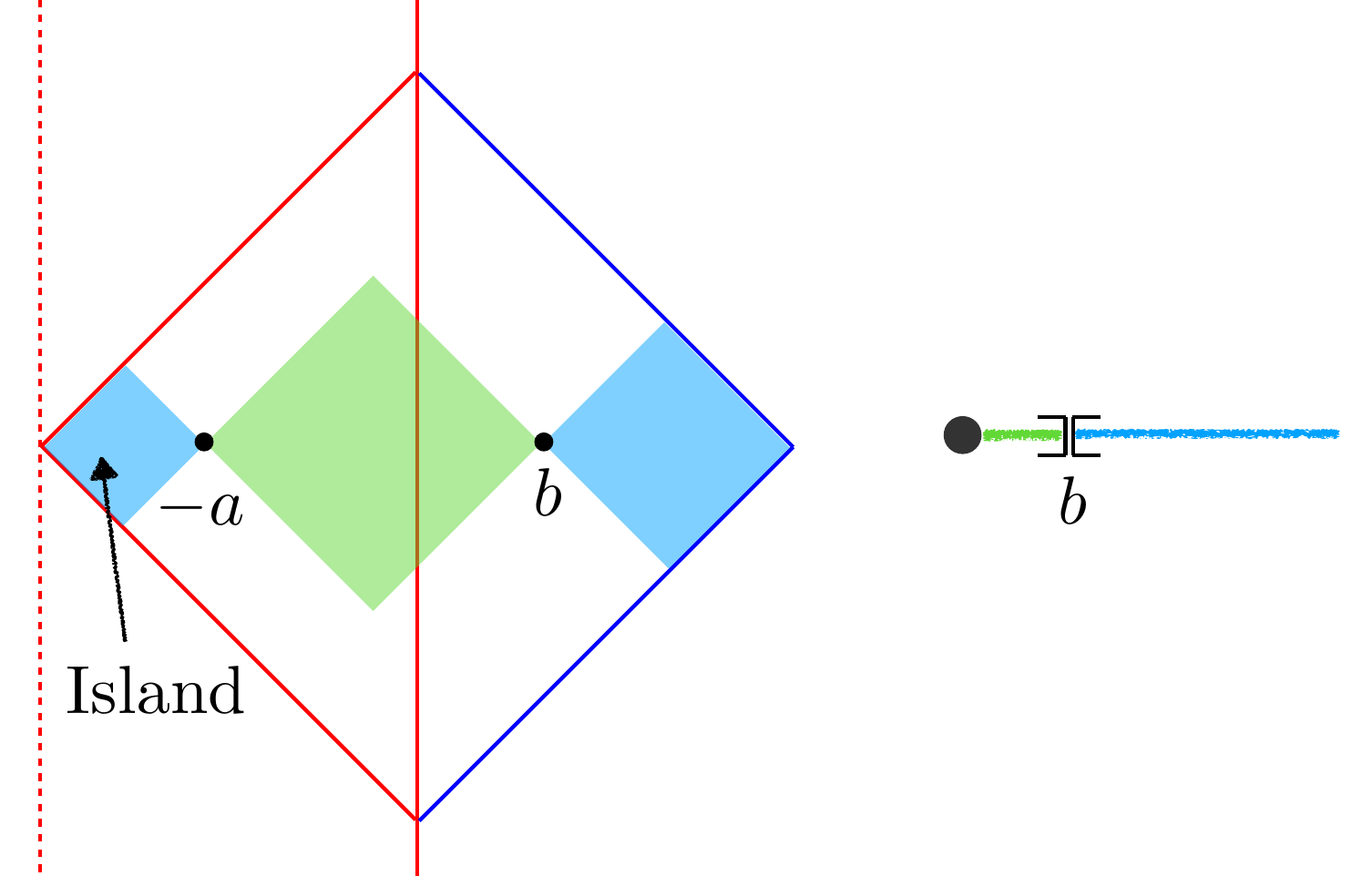}
    \end{center}
    \caption{On the left, in green, we see the entanglement wedge of the region corresponding to the interval ${\fbseries [0,b]}$. It is also shaded in green on the right. 
    Shaded in blue we see the interval $ {\fbseries [b,+\infty]}$ and its entanglement wedge which also includes the island $[-\infty,-a]$, which is {\it outside} the horizon.}
    \label{ZeroTRegion}
\end{figure}

We start by considering the spatial interval ${ \fbseries [ 0, b]}$ that contains what we could call the ``black hole'' degrees of freedom, as well as a piece of the bath. 
We denote by boldface the intervals in the full quantum mechanical picture. 
This interval is depicted in the right part of figure \ref{ZeroTRegion}.
In particular, the point ${  \fbseries 0}$ is where one can picture the quantum system that gives rise to the AdS$_2$ region (when it is decoupled from the bath).
  
In order to find the entanglement wedge, we guess that it is the causal domain of some interval $[-a,b]$ in the $y$ coordinates, see figure \ref{ZeroTRegion}.
Intervals without boldface are intervals in the effective field theory, which includes the gravitational system. 
Such intervals should not be confused with the ones in boldface. 
The generalized entropy functional is
\be \la{SgenTz}
S_{\text{gen}}(a) =  \phi_0+  { \phi_r \over a } + S_{\text{bulk}}  ~,\quad 
S_{\text{bulk}} = { c \over 6 } \log \left[ { (a + b)^2 \over a } \right] + {\rm constant}\, .
\ee
The bulk entanglement entropy in (\ref{SgenTz}) is basically that of an interval of length $a+b$ in a CFT$_2$.
The factor of $a$ in the denominator comes from the warp factor in the AdS$_2$ region of the metric \nref{AdSTm}. 
We have absorbed the bulk UV (proper distance) cutoff into $\phi_0$, and the bath UV cutoff into the constant.

Extremizing $S_{\text{gen}}(a)$ with respect to $a$, we obtain the location of the quantum extremal surface
\begin{align}
 \tilde a &= \frac{1}{2} \left ( 1 + \tilde b  + \sqrt{ 1 + 6 \tilde b + \tilde b^2 }  \right) ~,~~~~~~\tilde b := b\, {  c \over 6 \phi_r } ~,~~~~~~~
 \tilde a :=  a\, {  c \over 6 \phi_r } \, ,
\la{afromb}
\end{align}
where we defined dimensionless quantities $\tilde{a}$ and $\tilde{b}$.
Since the horizon is at $\sigma = -\infty$, and the value of $a$ in (\ref{afromb}) is finite, the quantum extremal surface is {\it outside} the horizon, see figure \ref{ZeroTRegion}. 

The formula \nref{afromb} contains a length scale $\phi_r/c$. 
This length scale sets the evaporation time of a black hole, if we started the black hole at a high temperature \cite{Engelsoy:2016xyb}.
This scale is shorter by a factor of $1/c$ than the time scale $\phi_r$ at which JT gravity (described by the Schwarzian mode) becomes strongly coupled.
Since we have approximated the gravitational part of the entropy by its classical contribution, we need that $a \ll \phi_r$.
So the present discussion is justified when $c\gg1$, so that $\phi_r/c \ll \phi_r$. 
Then there is a non-trivial range of values of $b$ where the quantum corrections to the Schwarzian can be neglected. 
However, for $b > \phi_r$, we need to include quantum corrections described by the strongly-coupled Schwarzian, and the analysis of this paper is not valid.\footnote{Unlike \cite{Almheiri:2019hni}, in this paper,  we do not demand that the CFT$_2$ has a three-dimensional holographic dual.}
Depending on the size of $b$ compared to $\phi_r/c$  we can approximate 
\nref{afromb} as 
\begin{align}
    a \approx \begin{cases}
        \frac{6 \phi_r}{c} \, \quad \text{if } b \ll \frac{\phi_r}{c}\\
        b \, \quad \quad \text{if } b \gg \frac{\phi_r}{c}\\
    \end{cases}\, .
\la{AppE}
\end{align}

If we assume that the state on ${\fbseries [0,+\infty]}$ is pure, this implies that the entanglement wedge of the complement region ${\fbseries [b,+\infty]}$ will contain 
the region $[-\infty, -a]$. 
This is an island that is outside the black hole horizon, see figure \ref{ZeroTRegion}. 
This calculation is essentially the same as the case we just considered: Since the global state of the quantum fields is pure, the bulk entropy of the region $[-\infty, -a] \cup [b,+\infty]$ is identical to that of its complement. 
The dilaton piece of $S_\text{gen}$ is shared by the two at the point $a$.

Had we not used the purity of ${\fbseries [0,+\infty]}$, we would naively consider the candidate quantum extremal surface for the bath region  ${\fbseries [b,+\infty]}$ to be just the surface which contains no island contribution.
The generalized entropy in this case would simply be the field theory entropy of the interval $[b,+\infty]$ in the vacuum, which has an IR divergence. 
Therefore this surface, while technically a quantum extremal surface, is never minimal.

Note that the entropy (\ref{SgenTz}) contains a term involving $\phi_0$, both for the 
region ${\fbseries [0,b]}$, as expected, but also for the bath region ${\fbseries [b, +\infty]}$, which might not have been expected.

Notice that a region of the geometry at some position $a$ corresponds to an energy scale of order $1/a$ or length scale of order $a$. 
So the fact that regions very close to the horizon (large $a$) corresponds to very long distances in the bath does not look too unreasonable. 
Let us discuss this more.

Consider the entanglement wedge for an interval $ {\fbseries [ b, b'] } $, with $b < b'$. 
We further take 
$ b \ll {\phi_r \over c } \ll b'$.  When there is no island the entanglement entropy is 
\be
 S_{\rm no-island} = { c \over 6 } \log (b-b')^2  \approx { c \over 6 } \log {b'}^2\, .
 \ee
When there is an island, say going between $-a'$ and $-a$, it is harder to compute the bulk entropy.
However, when $b'$ is large we can use an OPE argument, if the points $b'$ and $-a'$ are close. 
Naively $b'$ and $-a'$ are very far away, but they are both approaching the point at spatial infinity in Minkowski space.\footnote{ This closeness can be understood by viewing the two-dimensional conformally compactified Minkowski space as a patch wrapping the cylinder, where the two spatial infinites meet at a point.} 
Since the theory under consideration is conformal, we can do the OPE expansion (in the $n$-replica geometry). 
The leading term produces the entropy of two separate intervals $[-a',b']$ and $[-a,b]$.
For each of them we can use  \nref{afromb} to obtain 
\be \la{IslandTwo}
 S_{\rm island} = 2 \phi_0 + { c \over 6 } \left[ \log \left({ b' \phi_r \over c }\right)    + O(1) \right] \, ,
\ee
where we used the approximate expressions \nref{AppE} for $a$ and $a'$. 
If we want the island phase to dominate we need $S_{\rm island} < S_{\rm no-island}$, which implies that 
\be \la{LargeDis}
  \log\left(  { c  \, b' \over \phi_r } \right)   >   { 12 \phi_0 \over c } + O(1) \, .
\ee
Therefore we see that if $\phi_0/c$ is large, then $b'$ has to be exponentially large in order for the entanglement wedge of ${\fbseries [b,b']}$ to contain an island.\footnote{Recall that our discussion is valid when $b' < \phi_r$, so we need $c\gg 1$ and $\phi_0/c$ not too large.}

If we take the $b' \to \infty$ and $b \to 0$ limit of \nref{IslandTwo} we {\it do not} get the same entropy as in \nref{SgenTz}. 
This also happens in simpler situations.  
For example, we can compare the large size limit of a finite size interval in a CFT vs taking the whole line from the beginning (which would give zero entropy). 
The reason is related to the IR modes that continue contributing when we have a finite interval. 
In this particular case, these IR modes involve also $\phi_0$ and sense the presence of the boundary.  
  
To summarize, this discussion suggests that the degrees of freedom in the island are encoded in very long distance correlations in the exterior. 
We reached this conclusion by noting that the interval ${\fbseries [b ,b']}$ has to be very large for its entanglement wedge to contain the island.

\subsection{A consistency condition} 
\la{SecConsistency}

The fact that the island is outside the horizon looks surprising, because we can send a signal from the island to the boundary of AdS$_2$. 
This is not immediately a paradox since we are dealing with a coupled system where the time evolution mixes the degrees of freedom in ${\fbseries [0,b]} $ and ${\fbseries [ b, \infty]}$. 
This coupling certainly allows a region inside the entanglement wedge of ${\fbseries [b, \infty]}$ (such as the island) to send a signal to ${\fbseries [0, b]}$.

\begin{figure}[t!]
    \centering
    \begin{subfigure}[b]{0.34\textwidth}
        \centering
        \includegraphics[scale=0.39]{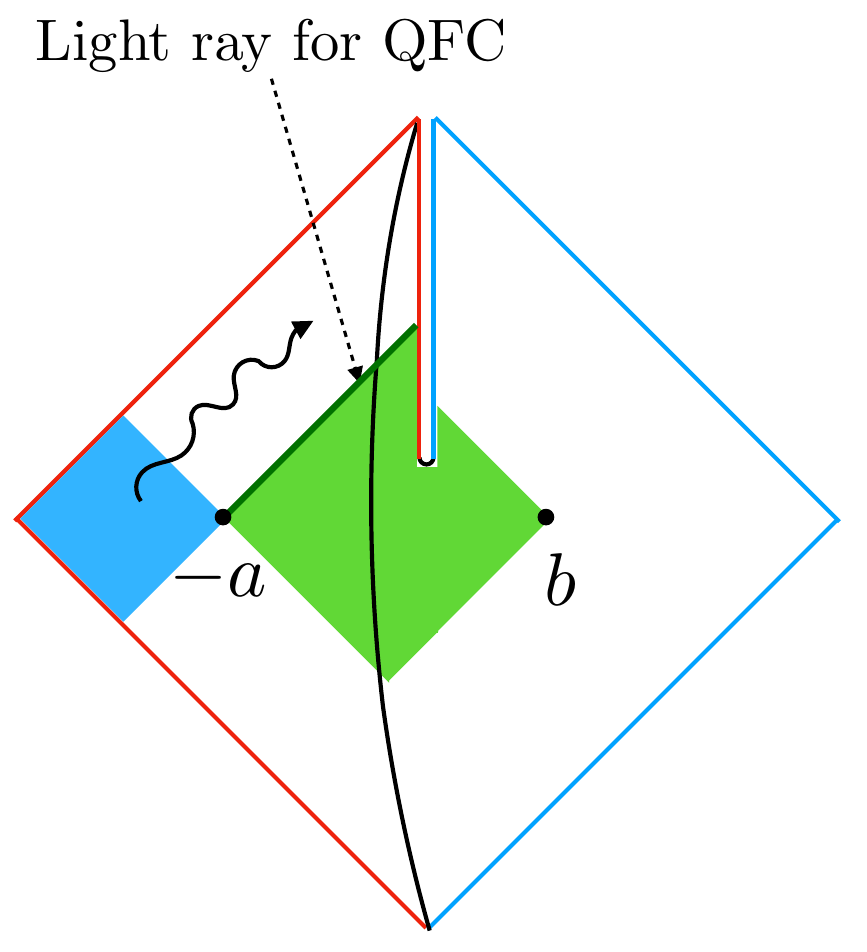}
        \caption{Ruled out by QFC}  
    \end{subfigure} 
    \begin{subfigure}[b]{0.34\textwidth}
        \centering
        \includegraphics[scale=0.39]{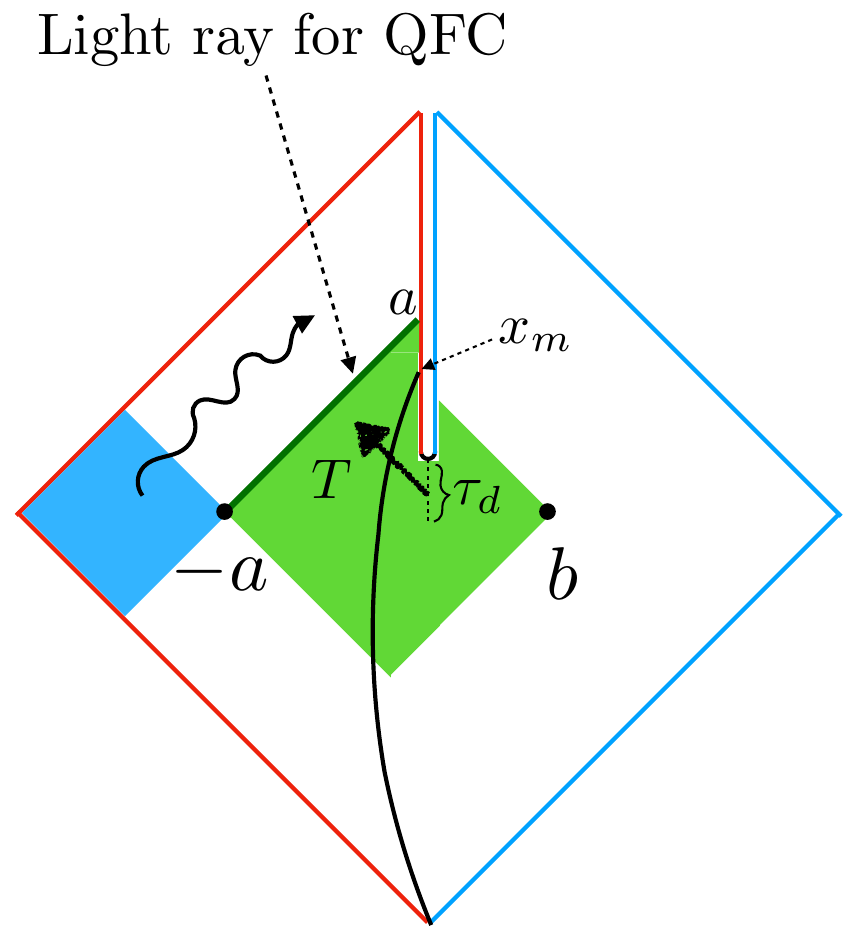}
        \caption{Required by QFC}
    \end{subfigure}
    \begin{subfigure}[b]{0.30\textwidth}
        \centering
        \includegraphics[scale=0.41]{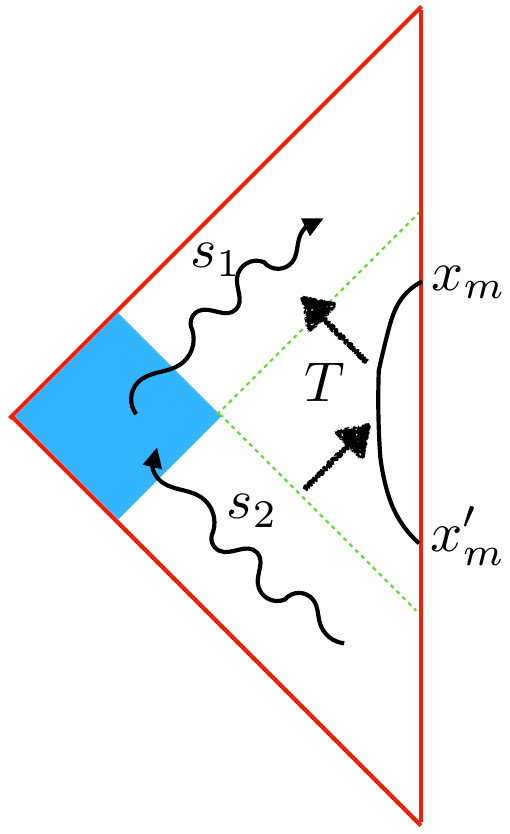}
        \caption{Decoupled backward evolution}
    \end{subfigure}
    \caption{
    Decoupling the black hole from the bath. 
    (a) Naively, a signal from the island can reach the AdS boundary. This is wrong. 
    (b) The decoupling process produces some energy that moves the horizon outwards. This means that the trajectory of the boundary particle (shown in solid black) reaches a maximum Poincar\'e time $x_m$. As long as $x_m < a$, we will have no contradiction since the signal will not be able to reach the physical boundary. This is guaranteed by the quantum focusing conjecture.
    (c) After we decouple, we can evolve the system backwards in time. Then we should also find that $x'_m >-a$ to avoid seeing the signal $s_2$.}
\label{Consistency}
\end{figure} 

The situation becomes more interesting if we decouple the black hole from the bath. 
More precisely, let us imagine that we decouple the localized degrees of freedom living at ${\fbseries 0}$ from the ones in the CFT. 
If we do this over a time shorter than ${  b}$, then no information from ${\fbseries [b ,\infty]}$ can reach the degrees of freedom at living at ${\fbseries 0}$. 

One might naively think that we get a Penrose diagram as in figure \ref{Consistency}(a), where the signal reaches the boundary.
However, this ignores the energy that is created by the decoupling process. 
This energy can raise the temperature of the black hole so that we end up with a diagram as in figure \ref{Consistency}(b), where the signal does not get to the physical boundary of the nearly-AdS$_2$ region. 
When we insert some energy, the map between the $x$-coordinates and the time of the quantum system is such that $x$ gets to a maximum value $x_m$. 
This map can be obtained in a variety of ways 
\cite{Engelsoy:2016xyb,Almheiri:2014cka,Maldacena:2016upp,Jensen:2016pah}.
To avoid a contradiction, we need that   $x_m < a $, as in figure  \ref{Consistency}(b). 
Furthermore, if we evolve the decoupled system backwards in time, we should \emph{not} receive any signal either, see figure \ref{Consistency}(c).

One expects that an amount of energy of order $c/b$ will be produced if we decouple the two systems over a time less than $b$. 
This amount of energy appears to be parametrically sufficient  to avoid a contradiction. 
The related fact that some energy is inevitably produced when we couple the system to a bath was discussed in \cite{AMS} and is used to avoid a potential contradiction also there.
This is also implied by the general argument in \cite{Engelhardt:2014gca}, applied after decoupling,  that the quantum extremal surface should be behind the horizon.\footnote{We thank Netta Engelhardt for discussions on this point, which we incorporated into v2 of this paper.} 
See also \cite{Wall:2012uf, Hubeny:2012wa} for the corresponding argument for classical extremal surfaces.
Below we go over a detailed argument for this special case, highlighting the assumptions.

We now show that the causality paradox is avoided if we assume the two-dimensional version of the quantum focusing conjecture \cite{Bousso:2015mna}
\be \la{QFCNaive}
 \nabla_+^2 (\phi + S)  \leq 0      
\ee
In the usual QFC, $S$ is the entropy of the bulk fields to the right of the light ray.
In our particular case, we will take $S$ to be the entropy of the interval that goes from a point along the light ray to the point $b$ in the bulk. 
In particular, at the start of the light ray, we have an interval from $[-a,b]$.
Since the decoupling process does not change the entropy of this interval, even after the decoupling the entropy of this interval continues to be the same, even though it is now broken into two pieces by the boundary.
In particular the quantum extremal surface continues to be a quantum extremal surface, so that $\nabla_+ (\phi +S) =0$ at the point $-a$.  
As we move along the light ray, the bulk region changes and the entropy (and the dilaton) will change.
However, the inequality \nref{QFCNaive}, together with the extremality condition at the starting point, imply  that the generalized entropy should decrease along the light ray that starts at the quantum extremal surface and moves towards the boundary.
This follows by integrating (\ref{QFCNaive}) twice with respect to an affine parameter along the light ray.
On the other hand, if this light ray were to reach the physical AdS$_2$ boundary, we would expect that the generalized entropy becomes very large: The dilaton  grows as we approach the boundary, but the entanglement entropy saturates.
This can be avoided if this light ray reaches the ``singularity,'' or the region that lies beyond the end of the physical AdS$_2$ boundary, as in figure \ref{Consistency}(b). 

Note that we only need to apply the QFC \nref{QFCNaive} to the state obtained after the decoupling procedure, on a space with a boundary. 
This state can be evolved forwards or backwards in time with the decoupled Hamiltonian to obtain the state along the light ray.\footnote{Here we implicitly assumed that at the decoupling time $\tau_d$, we have $x(\tau_d)<a$. 
The fact that this should be the case can be ensured again by using the QFC, but now taking $S$ to be the entropy of the region to the left of the light ray, together with the region to the right of $b$.}

In   two dimensions one can prove   \nref{QFCNaive} as follows\footnote{This argument was added in version 4. We thank  T. Faulkner, A. Levine and A. Wall for discussions.}.  
Using the equations of JT gravity  (\ref{QFCNaive}) becomes (see \cite{Strominger:2003br}) 
\be \la{Navex}
 - 2\pi \, T_{++} + \nabla^2_+ S \leq 0 
\ee
For a CFT$_2$, a general background simply amounts to a general Weyl factor for the metric. So  one expects that
  (\ref{Navex}) would follow from the conformal transformation properties of the flat space QNEC that was conjectured in \cite{Wall:2017blw} and proven in \cite{Bousso:2015wca,Balakrishnan:2017bjg,Ceyhan:2018zfg}
\be \la{QNEC}
- 2 \pi  \, T^{\text{flat}}_{++} + \partial_+^2 S^{\text{flat}} \leq 0 ~~~~~~~({\rm flat~ space, ~proven})\, .
 \ee
 In fact, using conformal transformations it is possible to prove the stronger inequality\footnote{This   was  proven  for holographic 2d theories in \cite{Koeller:2015qmn}. We thank A. Levine for pointing out this reference and for discussions on this topic.}
 \be \la{ConFl}
 - 2 \pi \, T^{\text{flat}}_{++} + \partial_+^2 S^{\text{flat}} +  { 6 \over c} \, ( \partial_+ S^{\text{flat}})^2 \leq 0 ~~~~~~~
\ee
This can be derived as follows  \cite{Wall:2011kb}. We consider the change of coordinates $x^\pm = f_\pm(y^\pm)$, $ds^2 = -dx^+dx^- = - f_+' f_-' dy^+ dy^-$. 
We now consider a generic state in the $y^\pm$ coordinates, with the entropy and stress tensor computed with a cutoff in the Minkowski $y^\pm$ coordinates, $ds^2 =-dy^+dy^-$. We now go to the $x^\pm$ coordinates and we do a Weyl transformation to the Minkowski $x^\pm $ metric. This changes the stress tensor and the entropy as 
\be 
{f_+'}^2 T_{x^+x^+} =   T_{y^+y^+} + { c \over 24 \pi  } \{ f_+ , y^+ \} ~,~~~~~S^x = S^y + { c \over 12 } \log( f_+' f_-' )
\ee
The QNEC in the $x$ coordinates implies 
\begin{eqnarray}
0 &\geq & f_+'^2 \left[ - 2 \pi T_{x^+ x^+} + \partial_{x^+}^2 S^x\right]  =f_+'^2 \left[ -2\pi T_{y^+ y^+} + \partial_{y+}^2 S^y - { f''_+ \over f'_+} \partial_{y^+} S^y - { c \over 24} { {f''_+}^2 \over {f'_+}^2 } \right] 
\cr 
0 &\geq &   -2\pi T_{y^+ y^+} + \partial_{y^+}^2 S^y  + { 6 \over c } ( \partial_{y^+} S^y )^2 - { 6 \over c }  \left( \partial_{y^+} S^y + { c \over 12 } { f''_+ \over f'_+ } \right)^2 \la{FinNQe}
\end{eqnarray}
This should hold for any function $f_+$ and any state. In particular, it should hold if we choose $f_+''$ so that the last square in \nref{FinNQe} is zero.  Therefore we conclude that \nref{ConFl} should hold. 
 
  We now recall the  
the relations between the stress tensor and the entropy for the metric $ ds^2 = e^{ 2 \rho } ds^2_{\rm flat}$ and $h$ any scalar function:
\be
   T_{++} =   T_{++}^{\text{flat}} + { c \over 12 \pi } [ \partial_+^2 \rho - ( \partial_+ \rho)^2 ] ~,~~~~ S = S^{\text{flat}} + { c \over 6 } \rho 
 ~,~~~~~~~\nabla^2_+ h  = \partial_+^2 h - 2 \partial_+ \rho ~ \partial_+ h \, .
\ee
These equations imply that, on a general background, (\ref{ConFl}) becomes
\be \la{Stronger}
 -2 \pi \,  T_{++} + \nabla^2_+ S + { 6 \over c }   ( \nabla_+ S )^2 \leq 0 \, .
\ee
This inequality is stronger than \nref{Navex}. So we succeeded in proving \nref{QFCNaive}.

Actually, \nref{QNEC} was derived for a single interval. If we have multiple intervals, then we can  argue as follows. Let us call the interval involved in the QNEC as the main interval and all other intervals as the ``others". We can then imagine putting a cutoff near the ends of all other intervals so that we can factorize the problem into type I algebras.  We can then unitarily map the region inside all other intervals into the inside of the main interval and join all the other outsides into a single outside. We leave the region near the endpoint of the main interval untouched so that the stress tensor remains the same. The entropy remains the same in this process. We can then use the proof for a single interval, which was valid for any state. In cases we have boundaries, we apply a similar argument. This is a rough argument which would be nice to make more rigorous.

 
Note that the QFC along the outwards  past-directed light ray ensures that the picture is also consistent if we evolve backwards in time, as in figure \ref{Consistency}(c).

Note that the QFC  (\ref{Stronger}) is also obeyed in the original coupled system, as in figure \ref{ZeroTRegion}.\footnote{We thank Douglas Stanford for raising this question.}
For example, we can check this on the right-moving light ray starting at the quantum extremal surface, that is, the point $-a$.
We take the region to be the spatial interval between $-a$ and $b$, and imagine moving the left end-point of this interval along the light ray at the top left edge of the green region of figure \ref{ZeroTRegion}.
In fact $\nabla_+^2 \phi = 0$ and also $\nabla_+^2 S + \frac{6}{c} (\nabla_+S)^2 =0$. 
One might then be confused because, in this case, the light ray can indeed get to the AdS$_2$ boundary, where the dilaton blows up. 
However, an important point is that since the right end-point has been fixed to $b$, we cannot move the left-end point beyond $x^+=b$.
In particular, this argument implies that $a>b$ since otherwise the AdS$_2$ boundary would intersect the light ray between $x^+=-a$ and $x^+=b$, and we would get a tension between the QFC and the dilaton growing to infinity at the AdS$_2$ boundary.
We can see that $a>b$ is obeyed in the expression (\ref{afromb}).

Notice that, even though the islands are outside the horizon of the coupled system, the islands always end up behind the horizon of the decoupled system, no matter how we decouple it. The details of the   decoupling process will determine the precise geometry and the actual position of the horizon. However, as discussed in \cite{Engelhardt:2014gca}, and reviewed  in this subsection,  in any unitary way of doing it, the QES always lies behind the actual horizon of the decoupled system.

\section{Nonzero temperature black hole coupled to a bath} 
\label{sec-finiteT}

\begin{figure}[t!]
    \begin{center}
    \includegraphics[scale=.45]{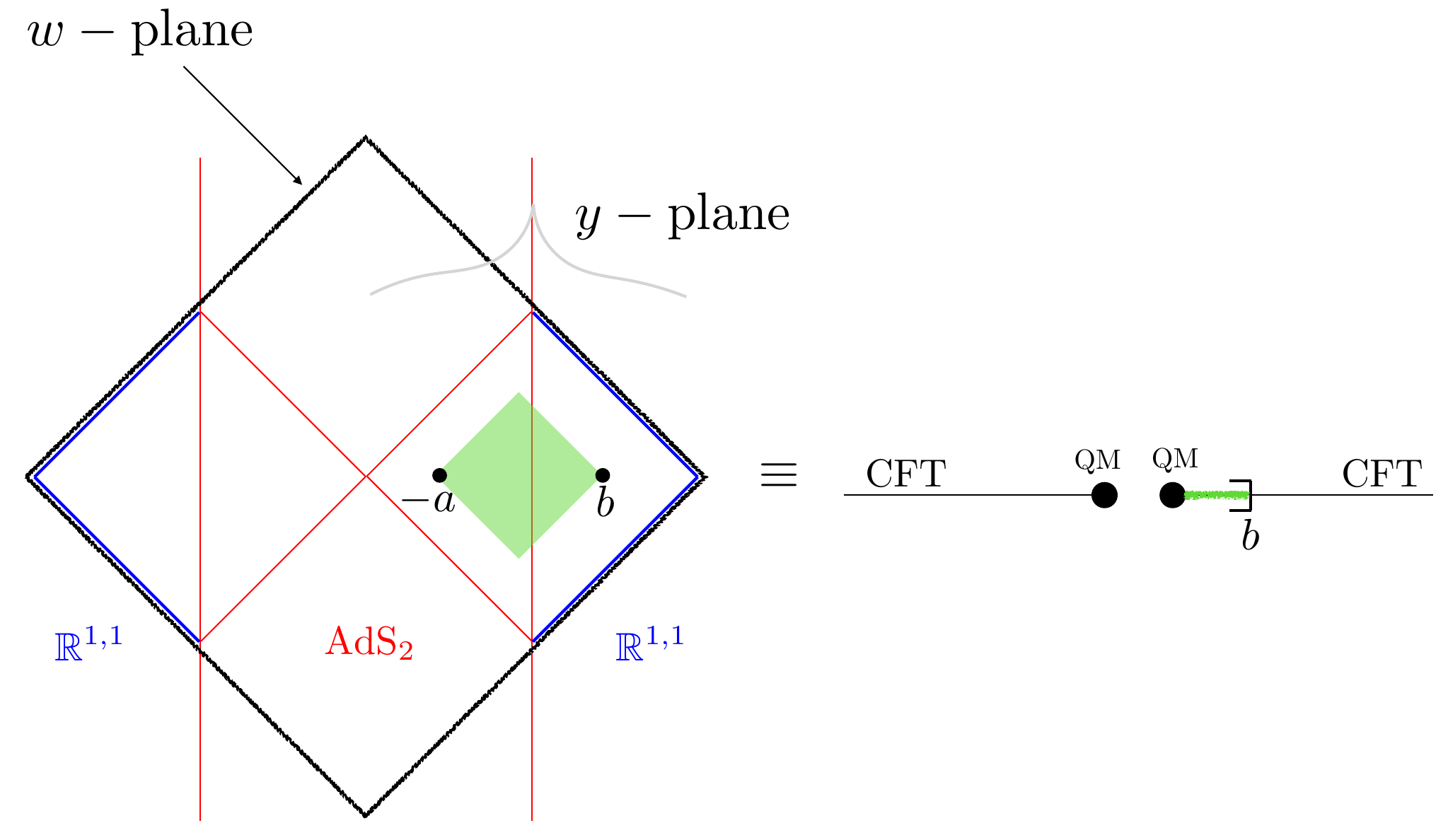} 
    \end{center}
    \caption{ A black hole in thermal equilibrium with a bath. The full Hartle-Hawking state is dual to  the thermofield double of the two quantum mechanical plus bath systems. We have two $y$-planes, one on the right and one on the left. Half of each $y$-plane belongs to the bath and the other half to the black hole exterior. The two $y$-planes fit into a $w$-plane where the state is the Minkowski vacuum for the quantum fields of the CFT.
     The $y$-planes are like Rindler wedges and the state looks thermal in the $y$-coordinates.}
    \label{ThermalPenrose}
\end{figure}

In this section, we consider black holes at nonzero temperature coupled to a bath.
More precisely, the setup can be thought of as having two copies of a black hole interacting with a bath, in the thermofield double state. 
This means that each black hole-plus-bath system is in a thermal state.
The state of the quantum fields on this background is the Hartle-Hawking state.
The geometry corresponds to an AdS eternal black hole (or AdS wormhole) that is connected to two half-spaces, one on each side, see figure \ref{ThermalPenrose}. 

The stress tensor in the bath half-space regions is the usual thermal one with constant energy density. 
The stress tensor in the eternal black hole or wormhole region is that of the usual AdS$_2$ vacuum. 
This implies, via \nref{StressT},  that the relation between the Poincar\'e coordinates $(x^+,x^-)$ and the right-exterior coordinates $(y^+,y^-)$ is 
\be \la{CoordR}
 x^\pm =  \tanh { \pi y^\pm \over \beta }\, .
\ee
The right black hole has the following metric and dilaton profile:
\be \la{FinTme}
ds^2 = - { 4 \pi^2 \over \beta^2}  {  d y^+ dy^- \over \sinh^2 { \pi \over \beta } ( y^- - y^+) } ~,~~~~~~~~
\phi = \phi_0 +  { 2\pi \phi_r \over \beta} { 1 \over \tanh { \pi \over \beta }(y^- - y^+ ) } \, .
\ee
The horizons are at $ y^- = + \infty$ and $y^+ = - \infty$. 
In the bath region, we define the coordinates $(y^+,y^-)$ as before.  
The whole $y$-plane describes one black hole exterior together with the bath.
A second $y$-plane describes the other side.\footnote{We have $x^\pm =- 1/\tanh( \pi y^\pm_L/\beta) $,  where $y^\pm_L$ are the coordinates of the left plane. The ones in \nref{CoordR} are the $y^\pm_R$ coordinates.} 

The computation of the bulk entanglement entropy is similar to that of a thermal state on the plane, except that we have to include the appropriate warp factor term from the metric \nref{FinTme}. 
In fact, the bulk entropy computation is simplest in the $(x^+,x^-)$ coordinates since the stress tensor vanishes and we have just the vacuum formulas. 
We then have to transform to $(y^+,y^-)$ coordinates and keep track of the warp factors and transformation of the UV cutoffs.

We consider an interval on the right side of the form ${ \fbseries [0, b]_R}$ that includes part of the right bath and the quantum mechanical degrees of freedom at ${\fbseries 0_R}$. 
We look for an entanglement wedge that consists of the interval $[-a,b]$, see figure \ref{ThermalPenrose}. 
Its generalized entropy is 
\be \la{SgenFT}
S_{\text{gen}}(a) =   { 2\pi \phi_r \over \beta} { 1 \over \tanh {2 \pi \over \beta }a } + { c \over 6 } \log {\sinh^2 { \pi (a+ b) \over \beta}
\over \sinh{ 2 \pi a \over \beta } } + {\rm constant} \, .
\ee
In the limit of large $\beta$, this reproduces \nref{SgenTz}.
Extremizing over $a$, we find    
\be
{ \sinh{ \pi  (a -b) \over \beta } \over \sinh {\pi (a+b) \over \beta } } = {12 \pi  \phi_r \over   c \beta } { 1 \over \sinh { 2 \pi  a \over \beta } } \, .
\la{EqFT}
\ee

The important point is that, again, the value of $a$ corresponds to a point outside the horizon, see figure \ref{ThermalPenrose}. 
This means that the region very near the horizon is not encoded in the degrees of freedom of the right boundary QM system. 
We find that $a\to \infty$ as $b\to \infty$, so that the quantum extremal surface approaches the black hole horizon.  
This is the statement that the entire right AdS$_2$ exterior is contained in the entire right system, ${\fbseries [ 0, \infty]_R}$, including the boundary QM system at ${\fbseries   0_R}$.

\begin{figure}[t!]
    \begin{center}
    \includegraphics[scale=.4]{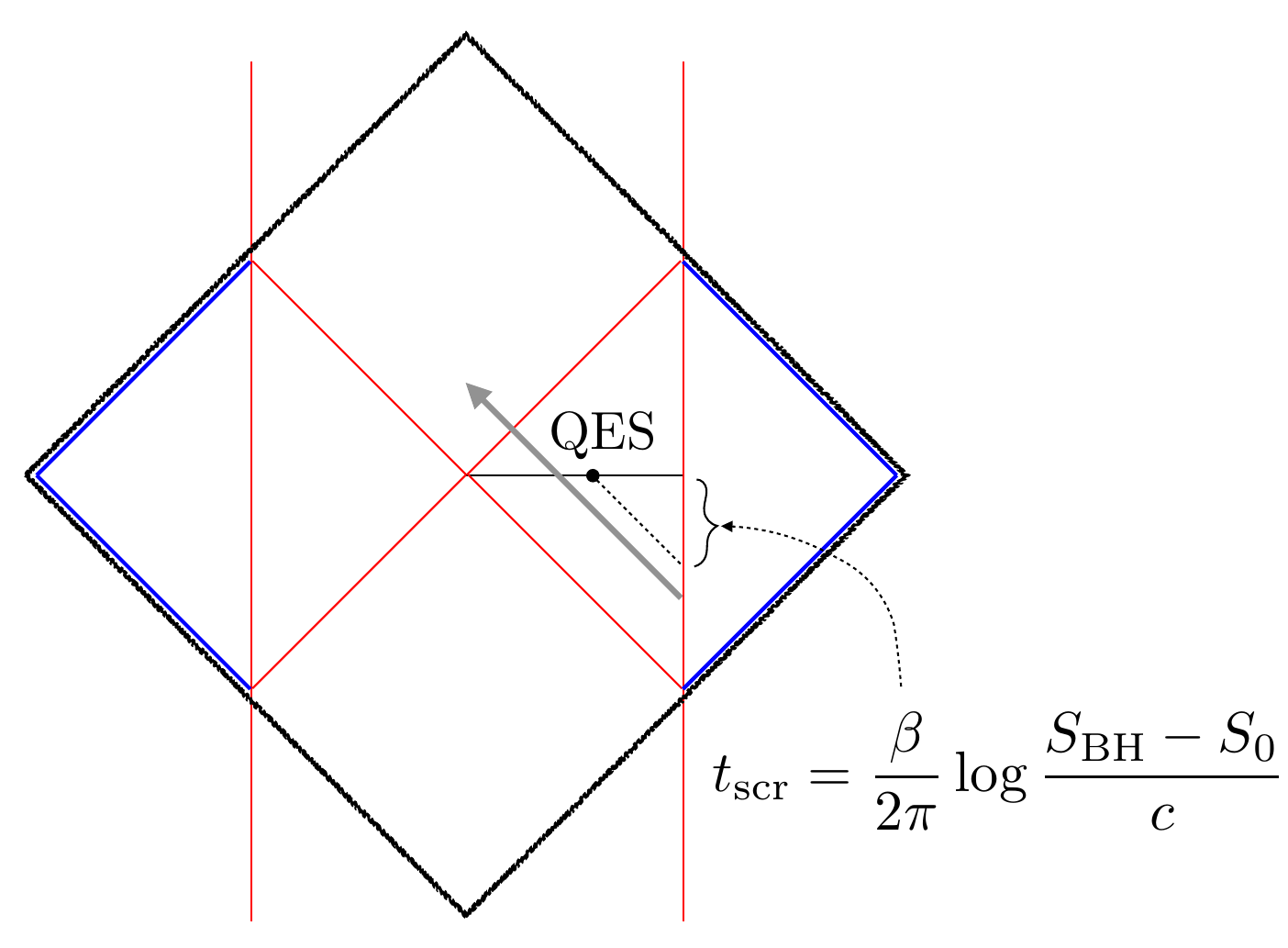}
    \end{center}
    \caption{A message inserted into the right boundary quantum mechanical system will exit its entanglement wedge after a scrambling time.}
    \label{hpisland}
\end{figure}
Note that \nref{EqFT} depends nontrivially on the ratio of the two length scales $\phi_r/c$ and $\beta$.
When $\phi_r/c \beta \ll 1 $, we get \nref{afromb}, whose extrapolation for large $b$ is also valid for   $b> \beta$.
In the opposite regime, when $ \phi_r/c \beta$ is large, the near-extremal black hole entropy is large compared to $c$ and the effect of Hawking radiation is a small perturbation.
In this limit, we get
\be \la{afrombFT}
 a \approx  b + { \beta \over 2 \pi } \log\left( { 24 \pi \phi_r \over c \beta} \right) ~,~~~~~~{\rm for }~~~~~{ \phi_r \over c \beta } \gg 1 \, .
 \ee
Notice that the island is causally connected to the right boundary. 
This means that if we deposit a message into the right boundary that is carried into the bulk by a particle falling into the black hole, this message will reach the island and thus will be lost to the right boundary QM system, see figure \ref{hpisland}.
From \eqref{afrombFT} we can see that in order to lose access to the message at $t = 0$, it needs to be thrown into the black hole roughly a scrambling time $t_\text{scr}$ in the past, where
\begin{align}
t_\mathrm{scr} =  { \beta \over 2 \pi } \log\left( { 24 \pi \phi_r \over c \beta} \right) + \ldots \, ,
\label{tscr}
\end{align} 
and we took $b \ll \beta \ll \phi_r/c$.
After time $t_\text{scr}$, the message should be reconstructable from the complementary system.
This is a manifestation of the Hayden-Preskill protocol \cite{Hayden:2007cs}.

Let us make some comments about the factor of $c$ in the denominator inside the log in (\ref{afrombFT}) and (\ref{tscr}).
Assuming $b \ll \beta \ll \phi_r/c$, the formula \nref{afrombFT} says that a simple signal sent at a time more than $t_{\text{scr}}$ in the past (from the time where we compute the island) will not belong to the right boundary QM degrees of freedom; see figure \ref{hpisland}.
This definition of the scrambling time, with a $c$ in the denominator inside the logarithm, might seem unconventional. 
But in fact, this definition agrees with \cite{Leichenauer:2014nxa}, after identifying the variables $E$,  $\delta E$ in \cite{Leichenauer:2014nxa}   with  $E \sim \phi_r/\beta^2$ and $\delta E \sim c/\beta$.
We can think of $t_\text{scr}$ as the time that has to elapse in order for the square of the commutator between the message and $c$ qubits of radiation to become of order one. 
If we had only one qubit of radiation, that time would be the conventional scrambling time (just the logarithm of the entropy). 
Since we have $c$ qubits, the commutator is $c$ times larger, and we need a shorter time, as in \nref{afrombFT} and (\ref{tscr}).

Finally, we comment that, once again, in the nonzero temperature situation, we could run into an inconsistency unless some energy is emitted when we decouple the system.
Again, the inconsistency is avoided by the use of the QFC \nref{QFCNaive}.

\subsection{Who claims the island?} 
\label{sec-claim}

Since the region on the right side between the horizon and the point $-a$, that is the interval $[-\infty,-a]_R$ in the $y$-coordinates, is not owned by the degrees of freedom localized at ${\fbseries [0,b]_R}$, we would like to understand who owns this region. 
More precisely, we want to ask from what region can we reconstruct the island, or extract the information that is contained there. 

It is clear that the island is in the entanglement wedge of  the region complementary to ${\fbseries [0,b]_R}$. 
The complementary region is the union ${\fbseries  
[-\infty,0]_L \cup [b,+\infty]_R }$. 
This is the whole left side, together with the outer interval on the right side. 
We are defining the coordinates on the left side so that negative values of $\sigma$ correspond to the bath, while positive values to the AdS$_2$ region.
So, in the bulk, the ``island" region $[-\infty, -a]_R$ is connected to the entire left side. 
Thus, more than an island, it is a peninsula that extends out from the left boundary.
In other words, the bulk entanglement wedge of  ${\fbseries  
[-\infty,0]_L \cup [b,+\infty]_R }$ is $[-\infty,\infty]_L \cup [-\infty,-a]_R \cup [ b, \infty]_R $. 
Note that the entanglement wedge of the whole left side, or ${\fbseries [-\infty,0]_L }$,  should be simply the whole left $y$-plane going up to the bifurcate horizon, or the interval $[-\infty,\infty]_L$. Therefore, by itself, it does not contain the island. 
  
\begin{figure}[t!]
    \centering
    \begin{subfigure}[b]{0.9\textwidth}
        \centering
        \includegraphics[scale=0.4]{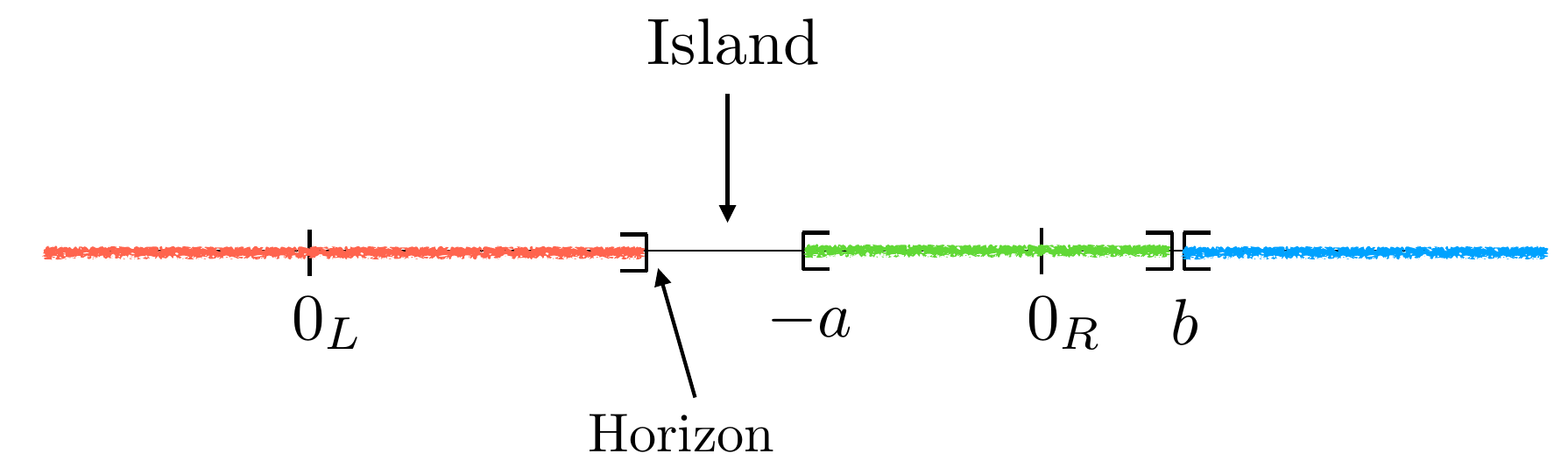}
        \caption{}  \vspace{0.3in}
    \end{subfigure} 
    \begin{subfigure}[b]{0.7\textwidth}
        \centering
        \includegraphics[scale=0.4]{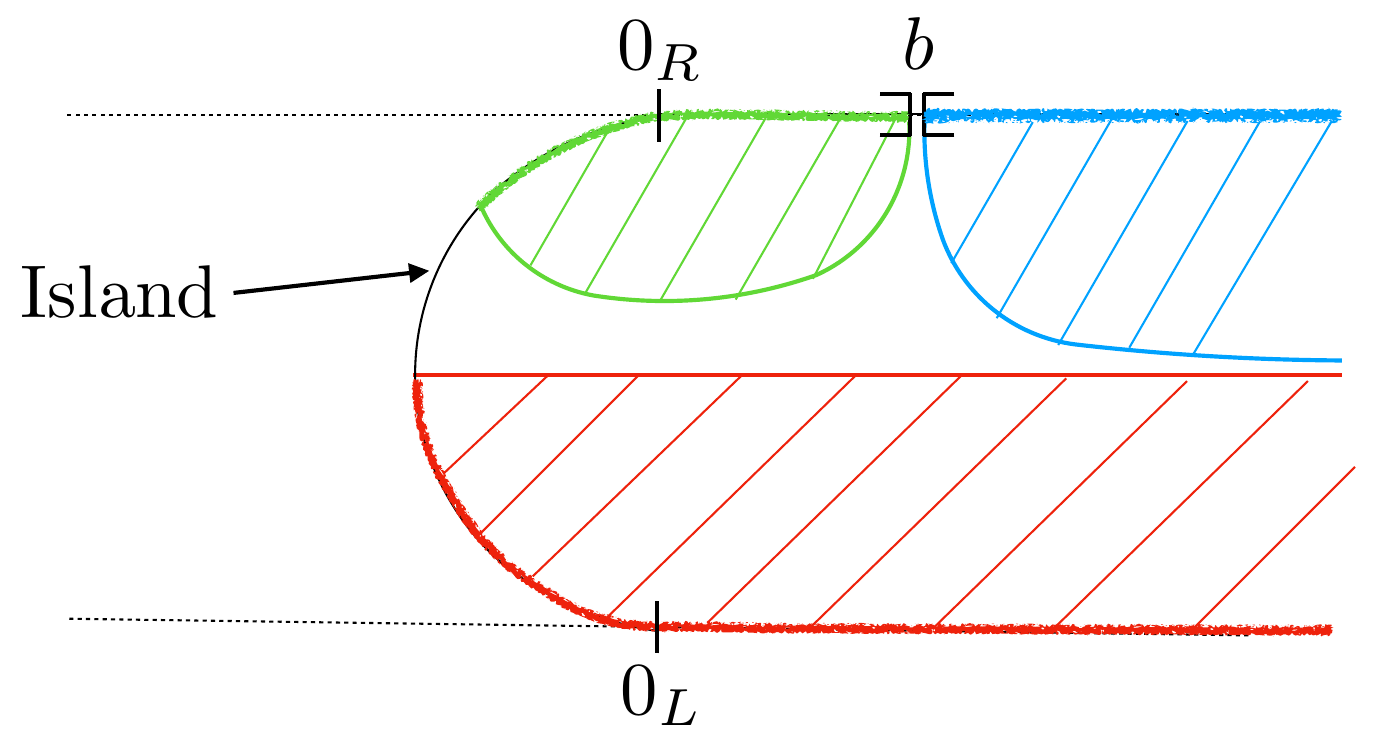}
        \caption{}
    \end{subfigure}
    \caption{In (a), we display a spatial slice of the eternal black hole coupled to two baths and the entanglement wedges for each of the three intervals ${\fbseries [-\infty,0]_L}$, ${\fbseries [0,b]_R}$, ${\fbseries [ b,\infty]_R}$. The island does not belong any single of them, but it does belong to any pair. In (b), we display the geometry when the matter CFT has a three-dimensional holographic dual, as in \cite{Almheiri:2019hni} (only a spatial section of the three-dimensional geometry is shown). }
\label{ThermalBulk}
\end{figure}   

We now consider the region ${\fbseries [b,+\infty]_R }$. We can argue that it cannot contain the island due to the following reason. If we have access to ${\fbseries [0,b]_R }$ and to the whole left side ${\fbseries [-\infty,0]_L }$, then we can 
extract the information that is in the island, at least the one sufficiently close to the horizon, by inserting a pair of operators in these two regions, as in the traversable wormhole/quantum teleportation discussion in \cite{Gao:2016bin}. 
In section \ref{Stranded} we will discuss this procedure in more detail. 
Therefore we conclude that the island cannot be in the entanglement wedge of the 
${\fbseries [b,+\infty]_R }$ region on its own.\footnote{
We thank Hong Zhe Chen and Rob Myers for pointing out an error in the argument we had in the first version of this paper. 
We also thank R. Bousso for suggesting the present argument.} 
 
The upshot is that this island does not belong either to the left side ${\fbseries [ -\infty,0 ]_L }$, or to the ${\fbseries [b,+\infty]_R }$ interval.
However, it does belong to the union of these two regions.
See figure \ref{ThermalBulk}(b) for a picture of the entanglement wedges for the case that the CFT$_2$ has a three-dimensioanl holographic dual.
In summary, the island $[-\infty,-a]_R$ belongs to any two of the three regions ${\fbseries [-\infty,0]_L}$, ${\fbseries [0,b]_R}$, ${\fbseries [ b,\infty]_R}$, but not to any single one of them. 
This is similar to the situation discussed in \cite{Almheiri:2014lwa}.
  
\subsection{Rescuing the information stranded in the island } 
\la{Stranded} 

From the previous discussion we expect that it should be possible to extract the information in the island if we have access to the union of ${\fbseries [-\infty,0]_L  \cup [b,\infty]_R}$.
In general, extracting information from an  island could be very difficult. 
However, for this particular system there is a relatively simple way to do it \cite{Gao:2016bin, Maldacena:2017axo}.
 
\begin{figure}[t!]
    \begin{center}
    \includegraphics[scale=.7]{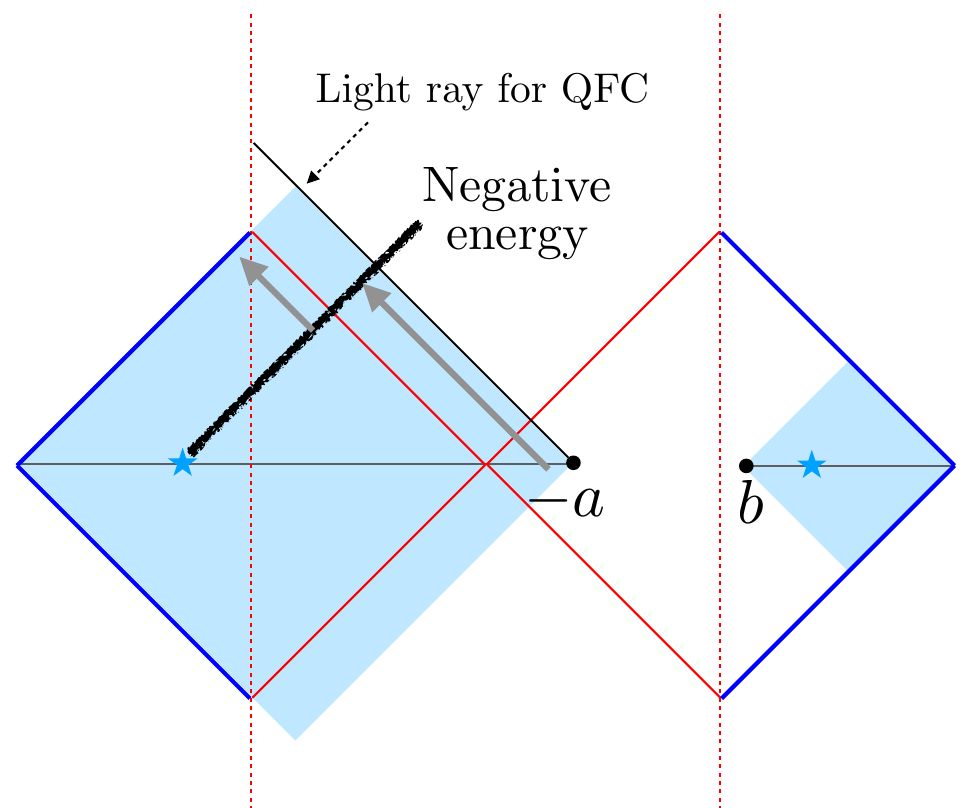}
    \end{center}
    \caption{Recovering the information stranded in the island through a quantum teleportation protocol. This involves measuring fields to the right of $b$ on the right side, and using that information to generate a negative energy pulse on the left side \cite{Gao:2016bin}. The natural evolution of the left side then extracts the information from the island (at least the information that is contained in purely left- or right-moving fields in the bulk).}
    \label{Recovering}
\end{figure}
  
For simplicity let us consider the case when $b \ll \beta \ll \phi_r/c$.
Then, from (\ref{CoordR}) and \nref{afrombFT} we see that the point $-a$ is such that its $x^+$-coordinate differs from that of the past horizon by
\be \la{xma}
 \Delta x^+(a) =    \tanh {  \pi (-a) \over \beta }  +1  \propto   { c \beta  \over \phi_r  } \, ,
\ee 
where we used \nref{afrombFT} to approximate $a$, with $b \ll \beta $. 
We can now follow the quantum teleportation ideas in \cite{Gao:2016bin} (see also \cite{Maldacena:2017axo}). 
By inserting a pair of operators, one near $\sigma = b $ in the right bath and the other somewhere on the left side, we can create a negative energy shock of magnitude $P_- = -i \partial_{x^-} \propto c$ that goes into the left black hole.\footnote{Note from (\ref{CoordR}) that $x^+$ and $x^-$ are dimensionless.}
Alternatively we can measure the operators on the right bath, send the information to the left side and apply a unitary that depends on the result.
Then the dynamics of two-dimensional gravity produces a shift in the horizon 
\be \la{NegShock}
 \Delta x^+_{\rm shock} \propto  { P_- \over \partial_r^2 \phi } \propto { c\beta  \over \phi_r   } 
 ~,~~~~~~~{\rm with} ~~~~~\partial_r^2 \phi \propto { \phi_r \over \beta  }  ~,~~~~~P_- \propto c \, .
\ee
The first formula is the general shift we have at the horizon of a black hole due to a pulse.   
The quantity $\partial_r^2 \phi$ is the Laplacian of the dilaton at the horizon and we used its expression in JT gravity, see \cite{Maldacena:2017axo} for more details.
Thus, we see that we can produce enough negative energy \nref{NegShock} to produce a shift of the right order of magnitude \nref{xma} that can pull the island into causal contact with the left system; see figure \ref{Recovering}.
We can also evolve the left system backwards in time on the left side, as was discussed for the zero temperature case in figure \ref{Consistency}(c). 

We should emphasize that this protocol involves both baths {\it plus} the left boundary quantum system. 
While the initial deformation is localized in the baths, evolving with the coupled Hamiltonian of the left boundary-plus-bath is crucial for propagating the negative energy into the bulk to pull out the island.

Note that we should not be able to overdo this and pull out {\it more} than the island. 
Therefore we have a sharp bound on how effective the teleportation process can be. 
Fortunately, by an argument similar to that of section \ref{SecConsistency}, the QFC ensures this. 
Namely, we can apply it along the  light ray that starts at the quantum extremal surface and goes to the left, and use the entropy of the spatial interval between a point on the light ray and  $b$; see figure 
\ref{Recovering}. 
Again we use that the point $-a$ is a quantum extremal surface so that the gradient of the generalized entropy vanishes there. 
As before, this implies that the light ray cannot hit the physical boundary of AdS$_2$ and come out on the left. 
One might be hesitant to apply the quantum focusing conjecture in a situation that we have a non-local operator. 
Note, however, that this operator is a unitary transformation that acts on the complement of the region $[-a,b]$. 
This implies that, at $-a$, we continue to have a quantum extremal surface. 
Furthermore, this unitary produces a new state and then we are simply applying the QFC for this new state.  
   
Note that this recovery protocol explains why the island extends outside the horizon, since it shows that the information in that region can be recovered without the knowledge of the right boundary QM system. 
In addition, the fact that the extent of the island is correct to be extractable, as discussed in \nref{NegShock}, shows that the factor of $c$ inside the logarithm in \nref{afrombFT} is correct. 

In summary, we have argued that we can extract some part of the island and that we cannot pull out more than the island.
However,  we did not show that we can pull out precisely the whole island. 
We leave this to the future.

\section{An information paradox for the eternal black hole} 
\label{sec:eternalinfo}

In this section, we want to point out that there is already a Hawking-like information paradox for the eternal black hole in equilibrium with a bath, as we considered above.\footnote{We thank Thomas Hartman and Douglas Stanford for discussion on this topic. This is different from the eternal black hole paradox discussed in \cite{Maldacena:2001kr}, which involved long time correlators for an eternal black hole which is {\it not} coupled to a bath. See also \cite{Papadodimas:2015xma} for a discussion of bulk local operators in the time-shifted thermofield double states.}
This is the thermofield double of the black hole and the bath together. 
There are two copies of the black hole and two copies of the bath. 
See figure \ref{ThermalPenrose}.  
A similar paradox was discussed in \cite{Mathur:2014dia}.

\begin{figure}[t!]
    \centering
    \begin{subfigure}[b]{0.48\textwidth}
        \centering
        \includegraphics[scale=0.32]{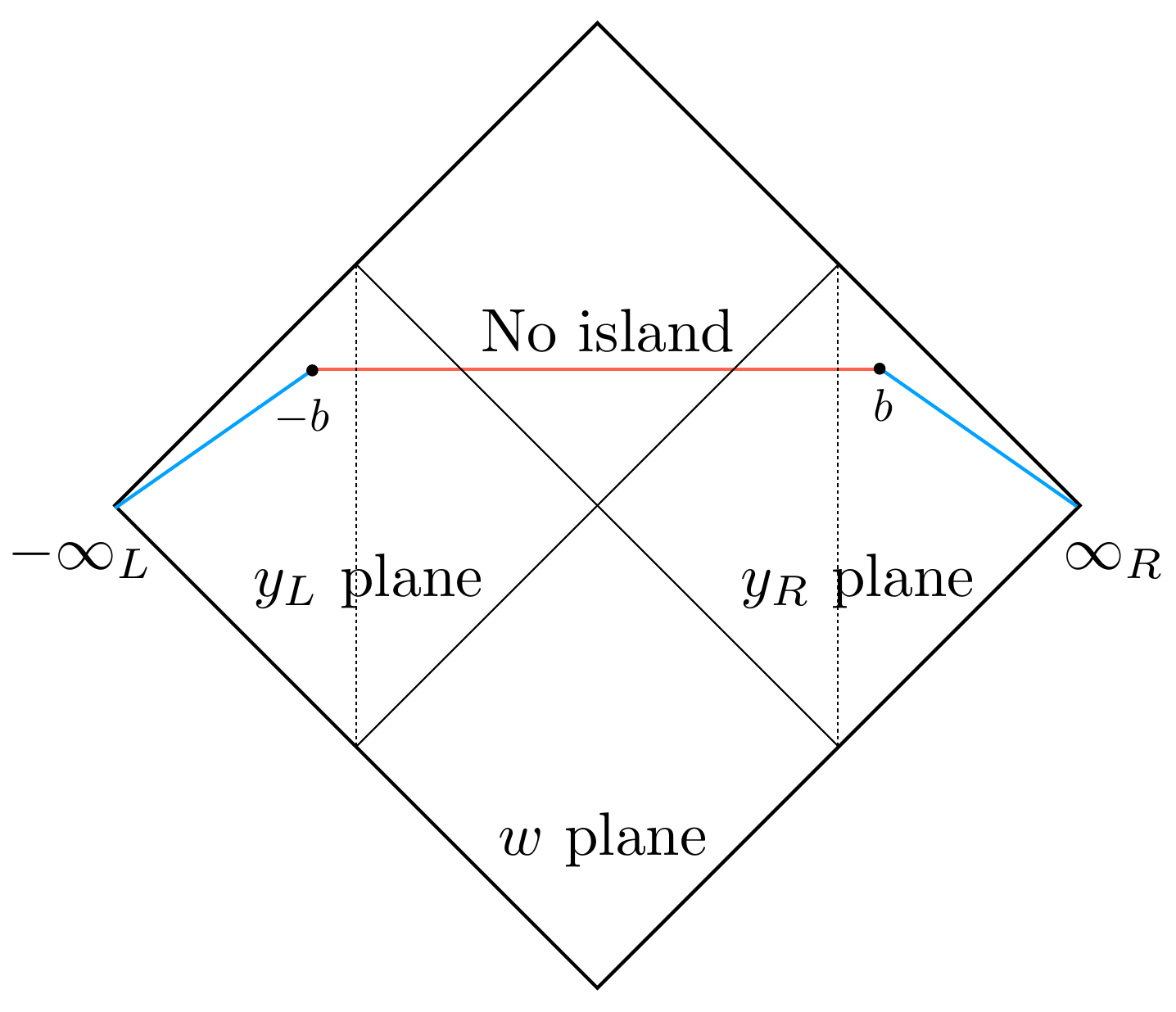}
        \caption{} 
    \end{subfigure}
    \begin{subfigure}[b]{0.48\textwidth}
        \centering
        \includegraphics[scale=0.32]{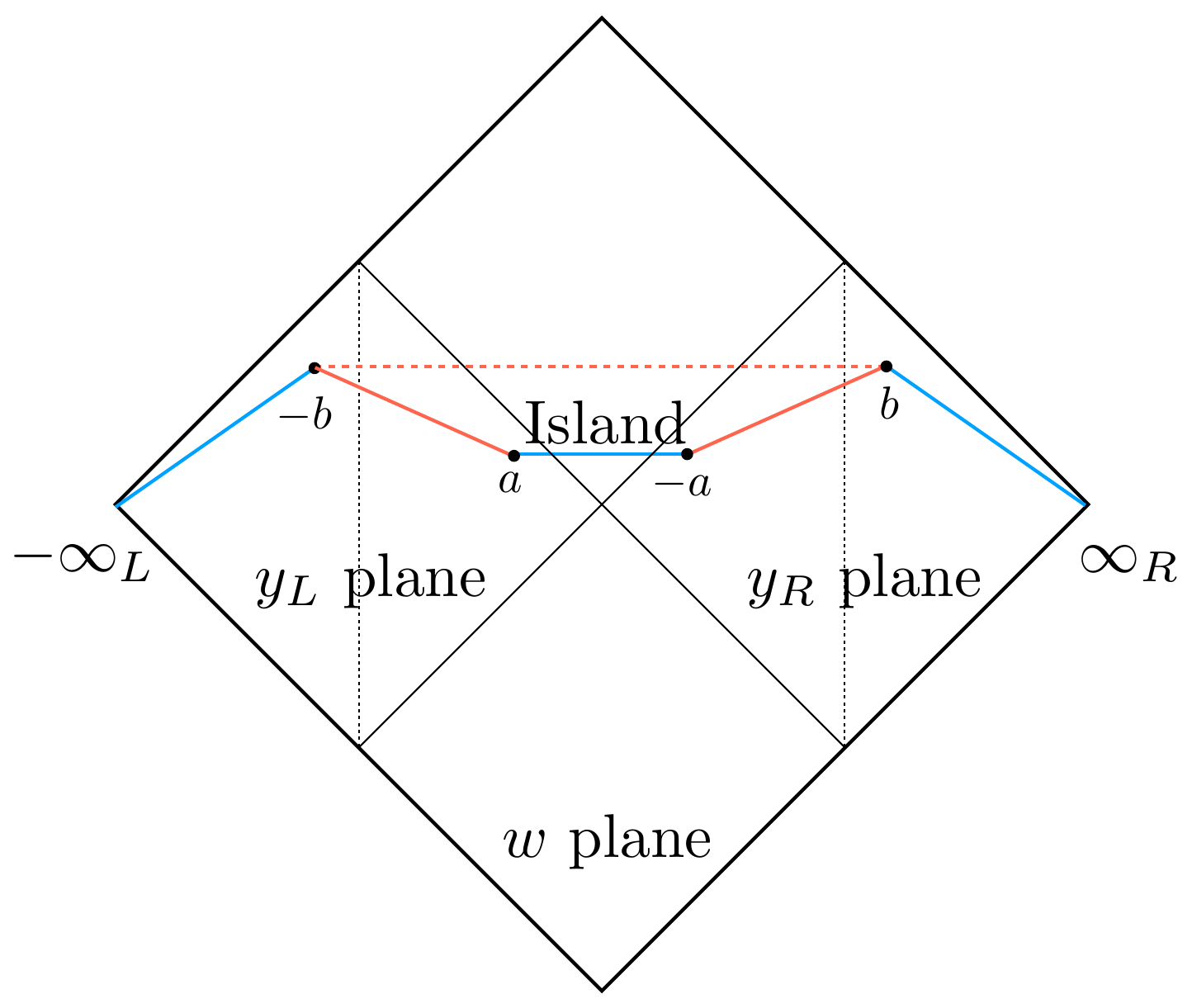}
        \caption{}
    \end{subfigure}
    \caption{An information paradox for the eternal black hole in the Hartle-Hawking state coupled to a bath. We consider the entropy of the two outside intervals ${\fbseries [-\infty,-b]_L \cup [b,\infty]_R}$, situated at the same time on the left and the right sides. 
    (a) The configuration with no island that gives an entropy that grows linearly with time.  
    (b) The configuration with an island. At late times, the distance between the two intervals $ [-b,a]_L $ and $[-a,b]_R$ is becoming very large, so we can do an OPE-like expansion to compute the bulk entropy, which gives the entropy of the two separate red intervals.}
\label{TwoIntervals}
\end{figure}

Let us consider the bath subregions ${ \fbseries  [-\infty ,-b ]_L}$ and ${\fbseries   [b, \infty]_R}$ both on the $t = 0$ slice. In this situation, the entanglement entropy can be computed as the entropy of two intervals in the thermofield double state and it is equal to 
\be \la{enNi}
 S = { c \over 3}  \log { \beta \over \pi } \, .
\ee
This is computed as follows. First we notice that we can view both   $y$-planes of figure \nref{TwoIntervals}  as  Rindler wedges of a single Minkowski space with coordinates $(w^+, w^-)$. These are related by
\be \la{Wpm}
w^\pm = \pm \exp\left( 
  \pm { 2 \pi y_R^\pm \over \beta }\right)~,~~~~~~~ 
  w^\pm = \mp \exp\left( 
  \mp { 2 \pi y_L^\pm \over \beta }\right)\, .
  \ee
The state in the $w^\pm$ plane is just the vacuum. 
So the entropy is 
\be
  S = { c \over 6 } \log\left( 
    { w_{12}^+ w_{12}^- \over \sqrt{ \partial_{y^+ } w_1^+ \partial_{y^- } w_1^-
    \partial_{y^+ } w_2^+ \partial_{y^- } w_2^- } } 
   \right)  \, .
\la{WplaEn}
\ee
The denominators arise from the change in the cutoff between the $y$-plane and the $w$-plane.
This formula is appropriate when the endpoints are in the bath region (in the black hole region we should take into account the warp factor). 
Substituting $t=0$, $\sigma_{L} =  - b$ and $\sigma_R = b$ into \nref{Wpm} and (\ref{WplaEn}), we get \nref{enNi}.
  
The entropy \nref{enNi} is just that of the interval $[-b_L, b_R]$ of the quantum fields through the wormhole. 
This is a finite interval in the $w$ plane, see figure \ref{TwoIntervals}.
The addition of islands is not expected to decrease the entropy, for reasons similar to the ones given in section \ref{sec-claim}. 
So, we have no island and the entanglement wedge is simply the union of $[-\infty,-b]_L \cup [b, \infty]_R $.
Equation \nref{enNi} also gives the entropy of the two boundary quantum systems together when we consider the union of the intervals ${ \fbseries   [- b,0 ]_L \cup [0, b]_R}$. 

So far we have taken the intervals to lie on the $t=0$ slice on both sides. 
The situation becomes more interesting if we take the intervals to lie on a non-zero timeslice $t$. 
Moving the right side forwards in time and the left one backwards in time is an isometry.
Using this isometry we can make the left and right time equal if we had chosen to move them by an unequal amount.
So we will move the  the left and right times forwards in time on both sides, see figure \ref{TwoIntervals}. 
So now the values of the  coordinates of the left and right points are
\be
y_L^\pm = t \mp b ~,~~~y_R^\pm = t \pm b   \, .
\ee
Inserting these into \nref{Wpm} and (\ref{WplaEn}), we find that the entropy becomes   
\begin{align}
S &= {c \over 3} \log \left[{ \pi \over \beta} \,  {  \cosh\left( { 2 \pi t \over \beta}  \right) } \right]\longrightarrow {2 \pi \over 3} \, c \, {  t \over \beta }  + \ldots  ~~~~ {\rm for } ~~ t\gg \beta
\, . 
\label{trivialQES}
\end{align}
In (\ref{trivialQES}), we have also indicated the behavior for late times.
The dots are just constants that are independent of time.\footnote{From the bulk field theory point of view we have two copies of the $y$-plane in the thermofield double state and we are considering a region consisting of essentially a half space on each of the $y$-planes. This type of region, and this  linear growth in entropy,  was previously discussed  in \cite{Hartman:2013qma}.}

After a few thermal times we get linear growth with a rate governed by the temperature of the black hole and the matter central charge. 
This growth captures the exchange of Hawking radiation between the bath and the black hole. 
In other words, the bath is sending excitations into the black hole and the black hole seems to give them back in a way that is uncorrelated to what is falling in.
We are feeding the black hole and we are getting out thermal radiation, so the entropy growth in \nref{trivialQES} is a version of the information paradox; see also \cite{Mathur:2014dia}. 

It would be a paradox only if it went on for a long enough time. 
Initially the two boundaries are entangled with each other, and their entropy is also given by \nref{enNi} which is relatively small. 
Their evolution coupled to the bath is disentangling them from each other and entangling them to the bath. 
However, the maximal entropy we should be able to generate is 
\be
S_{\text{max}} = 2 S_{\text{BH}} = 2 \left( \phi_0 + { 2 \pi \phi_r \over \beta } \right) \, ,
\ee
which is the coarse grained entropy of the two black holes.

This linear growth in the entropy is reminiscent of Hawking's original information paradox as it leads to the `overfilling' of the black hole, whereby the fine-grained entropy exceeds the coarse-grained entropy of the wormhole.
In the present context, this is in tension with the finite number of states of the boundary QM systems.
In order to solve the paradox, we should find an effect that modifies the linear growth in \nref{trivialQES}.

\begin{figure}[t!]
    \begin{center}
    \includegraphics[scale=.6]{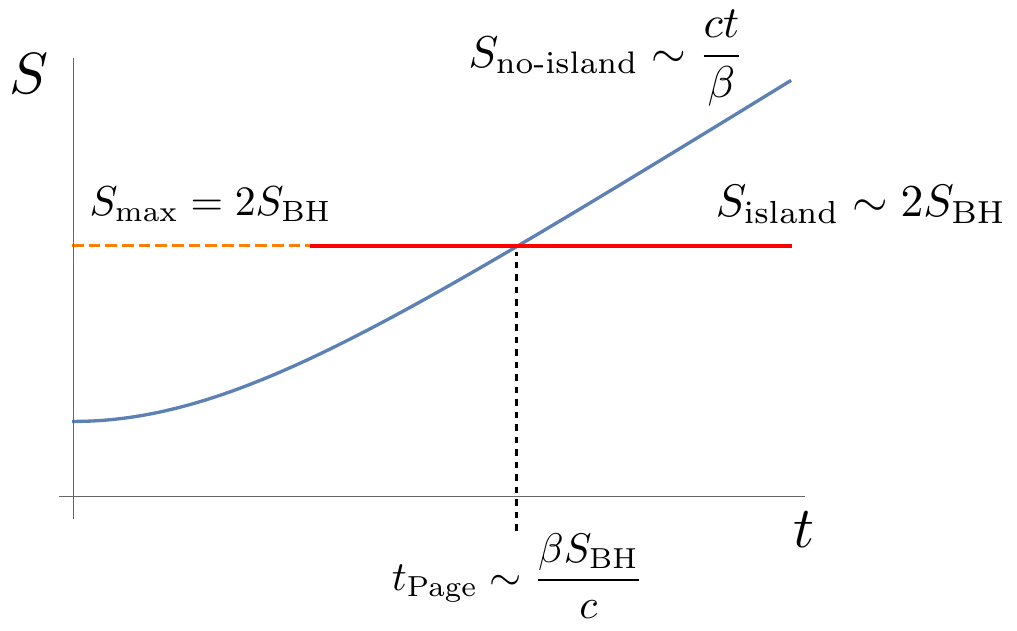}
    \end{center}
    \caption{The blue curve is the naive fine-grained entropy of the Hawking radiation, represented by the interval ${\fbseries [-\infty,-b]_L \cup [b,\infty]_R}$. It grows linearly after a few thermal times. Thus, it ultimately exceeds twice the Bekenstein-Hawking entropy of the black hole. This is a manifestation of the information paradox, which sets in at the Page time $t_{\text{Page}} \sim \beta S_{\text{BH}}/c$. The paradox is resolved by the orange curve, which is the entropy computed using the nontrivial quantum extremal surface with an island.}
    \label{entropycurve}
\end{figure}

For that purpose, we can try to include an island of the form $[-a,\infty]_L \cup [-\infty,-a]_R$. 
Now, in order to compute the entropy of the bulk quantum fields with this island included we have to consider a problem with two intervals, see figure \ref{TwoIntervals}(b).
In general this is difficult since such an entropy is a function of the cross-ratio of the four endpoints of the two intervals.
The actual value of the cross-ratio will come from extremizing the generalized entropy. 
However, our guess is that the value of $a$ is the same as in \nref{EqFT}. 
Let us check that this guess is correct.

The configuration of points at late times is as shown in figure \ref{TwoIntervals}(b).
An important observation is that, for this configuration of points, the cross-ratio of the four points is going to zero (or one or infinity, depending on how we define it).
Each interval has time-independent fixed proper size in the $w$-plane since they are related by a boost, while the distance between them is increasing.  
Therefore we can do an OPE in the channel where the points $-a$ and $b$ are close.  
(This OPE is done for the $n$-replica geometry.)  
The leading contribution in this OPE is from the identity operator. 
This leading contribution gives an entropy for the pair of intervals which is twice that of a single interval, given by the second term in \nref{SgenFT}. 
And the full generalized entropy for the pair is given by twice the left hand side of \nref{SgenFT}, whose extremization reproduces \nref{EqFT}. 
Therefore we have checked, a posteriori, that our assumption regarding the value of $a$ was correct.
The value of the entropy for this configuration is then given by 
\be \la{SIsla}
S = 2 \left( \phi_0 + {2 \pi \phi_r \over \beta} \right) + {\rm small}  \approx 2 S_{BH} ~,~~~~~~{ \rm for} ~~~ { \phi_r \over c\beta } \gg 1 \, .
\ee
Here $S_{\text{BH}}$ is Bekenstein-Hawking entropy computed using a single horizon.

The derivation of this second quantum extremal surface is valid when $t\gg \beta$, so that our OPE expansion is correct. 
On the other hand, the full entropy is given by the minimum of the no-island expression \nref{trivialQES} and the island expression \nref{SIsla}. 
We conclude that this entropic information paradox is resolved, in the sense that the entropies, computed using the RT/HRT/EW prescription \cite{Ryu:2006bv,Hubeny:2007xt,Engelhardt:2014gca},  give a result which is consistent with the expectations from unitarity; see figure \ref{entropycurve}.

Of course, the resolution is conceptually the same as what was discussed for an evaporating black holes \cite{Penington:2019npb,Almheiri:2019psf}.
An interesting aspect of this paradox and its resolution is that we do not need to consider any gravitational backreaction for the black hole. There is no change to the geometry of the black hole. What is happening is that the thermal bath is feeding the black hole at the same rate that it is evaporating. 
We should emphasize that this setup where we consider a pair of black holes and baths starting in a pure state is conceptually similar to a black hole that starts from collapse and then evaporates. 
It is simply that, in this eternal black hole case, the black hole solution is simpler to find; it is just the standard solution. 
All that changes is which part of the geometry is the entanglement wedge.

It is interesting to note that this setup also has an AMPS paradox \cite{Almheiri:2012rt} at times after the Page time. 
The basic argument is that if an outgoing Hawking particle is entangled with its partner behind the horizon, then it must raise the entanglement entropy of the bath, which would then conflict with the saturation of the entanglement entropy after the Page time (see figure \ref{entropycurve}). 
It was suggested that the black hole develops a firewall at the horizon at those times. 
However, due to the boost symmetry of the problem, one would have to conclude that a firewall is present along the entire bifurcate horizon at all times. 
Here, this conclusion is avoided due to the islands that realize $\text{ER} = \text{EPR}$ \cite{Maldacena:2013xja} (see also \cite{Susskind:2012uw, Papadodimas:2012aq}), where the build up of entanglement leads to the interior modes being included inside the entanglement wedge of the bath.

This paradox (and the resolution) is also present for higher-dimensional black holes in equilibrium with a bath.
As long as we are at finite temperature and the radiation is an infinite system, or a system with entropy much larger than the black hole, we will have the same paradox and we expect the same resolution. 
The precise location of the quantum extremal surface can change slightly, but conceptually we expect the same answer. 
In particular, we also expect the island to extend outside the horizon. 
We could even consider the flat space Schwarzschild black hole in the Hartle-Hawking state, as long as we neglect the backreaction of radiation on the flat-space region. 

\section{``Phenomenological" applications} 
\label{sec:pheno}

\subsection{Four dimensional near extremal black holes } 
\label{sec:fourd}
 
In this subsection we describe a situation where the considerations of this paper apply for certain four-dimensional black holes.
We consider near-extremal charged black holes whose low-energy dynamics is well described by the Jackiw-Teitelboim theory \cite{Almheiri:2014cka,Nayak:2018qej, Sachdev:2019bjn}. 
For a standard magnetically-charged black hole the relation between the parameters is
\cite{Maldacena:2018gjk} 
\be \la{Soluse}
\phi_0 = {\pi r_e^2 \over G_N} ~,~~~~~~
 \phi_r = { 2 \pi r_e^3 \over G_N}  ~,~~~~~~ {\rm with} ~~~~r_e^2 := { \pi q^2 G_N \over e^2 } ~,
\ee
where $q$ is the quantized (integer) magnetic charge and $e$ is the coupling for the $U(1)$ gauge field. 
Furthermore, if we have a massless charged fermion, then in the near-horizon region we get $q$ effectively massless two-dimensional fermions \cite{Maldacena:2018gjk}. 
 
This gives a setup similar to what was described above. 
We can ignore the effects of quantum gravity in the flat-space region far from the black hole and approximate it by a set of $q$ massless fermions following the magnetic field lines \cite{Maldacena:2018gjk}.
Deep inside the AdS$_2 \times S^2$ throat we have a situation described by Jackiw-Teitelboim theory coupled to matter with $c=q$. 
This is large if $q$ is large. 
The rest of the fields can be ignored for the purposes of computing the bulk entropies because there are not so many of them.

We encounter this situation for a magnetically-charged black hole in the Standard Model of particle physics whose size is small enough that we can approximate the fermions as massless. 
In this case the $U(1)$ is the weak hypercharge.   
Demanding that $r_e \ll 1/{\rm TeV}$,  we get $q \ll 10^{15}$. 
The effective two-dimensional central charge in this case becomes \be \la{centralch}
 c = 54 q\, .
\ee   
Thus the discussion of this paper precisely applies to this case, including numerical factors in all formulas, after using \nref{Soluse} and \nref{centralch}.\footnote{In particular, note that the length scale $\phi_r/c$, which here was setting the position of the island for zero temperature and small $b$ (\ref{AppE}), also appeared in \cite{Maldacena:2018gjk} as the effective ``length'' of the wormhole, defined as the time it takes to go through the wormhole from the outside point of view.}

\subsection{Comment on the transplanckian problem in Hawking radiation and cosmology}

The phenomenon of Hawking radiation and the creation of particles by the expansion of the universe are very closely related. 
In suitable coordinates, we can think of Hawking radiation as due to the expansion of space in the near horizon region. 
So if we take a given Hawking mode emitted at late time and we extrapolate it backwards in time by more than the scrambling time, ${ \beta \over 2 \pi } \log S_{\text{BH}}$, we find that it would have transplanckian energies. 
Normally, we say that this is not a problem because we are simply talking about the vacuum. 
Nevertheless, it has often been speculated that the derivation of Hawking radiation would fail once we reach the scrambling time.
This is a shorter time scale compared to the time scales at which we need corrections in the Page curve due to unitarity. 
Here, following \cite{Penington:2019npb,Almheiri:2019psf}, we have argued that the formation of islands avoids the problem related to the Page curve.  
We interpret this result as suggesting that one can get results consistent with unitarity without having to make the bulk theory break down at the scrambling time.
In other words, both the bulk and boundary points of view are correct, but we need to relate them in a new way.

In cosmology, one can worry about a similar transplanckian problem. 
Could a primordial fluctuations be extrapolated backwards in time so as to become transkpanckian?
The analog of the scrambling time in an approximately de Sitter space is 
$H^{-1}\log ( { M_\text{p}^2 / H^2})$. 
It has been recently conjectured in \cite{Bedroya:2019snp} (see also e.g. \cite{Martin:2000xs, Brandenberger:2012aj,Jacobson:1991gr}) that quantum gravity implies that we cannot have an approximately de Sitter universe for more than this time.
We view the absence of known paradoxes for the black holes  at these time scales as a reason not to worry about cosmology at these time scales.

\section{Discussion}
\label{sec-discussion}

In this paper we discussed various aspects of quantum extremal surfaces and entanglement wedges of black holes coupled to a bath.
We considered simple non-evaporating situations where the black hole is in equilibrium with the bath.
The geometries are simple and easy to describe explicitly.

The first interesting surprise is that the quantum extremal surface for a region that includes the boundary QM degrees of freedom and a little bit of the bath is {\it outside} the horizon. 
This is true for zero as well as nonzero temperature.
If one assumes entanglement wedge reconstruction, then one should conclude that the region between the horizon and the quantum extremal surface should be recoverable from the bath degrees of freedom. 
While it is not clear how to do this for general islands, we have shown how to recover the information from the nonzero temperature island by using the Gao-Jafferis-Wall teleportation protocol \cite{Gao:2016bin}. 
It might seem strange that the coupling to the bath effectively lets us ``reach'' into the near-horizon region (we are also using the system the black hole is entangled with).

The fact that the quantum extremal surface is outside the horizon makes us worry that we could run into causality paradoxes if we decouple the black hole from the bath. 
Physically, we are protected by the fact that the decoupling process will inevitably produce a suitable energy flux into the black hole. 
This energy flux moves the horizon outwards, and the quantum extremal surface then lies behind the horizon.
The quantum focusing conjecture (QFC) \cite{Bousso:2015mna} ensures that enough energy is produced.
In two dimensions, the QFC follows from a stronger version (\ref{ConFl}) of the quantum null energy condition in flat space.
This stronger version was conjectured in \cite{Wall:2011kb} and proven for holographic theories in \cite{Koeller:2015qmn}. 
So, even though the islands are outside the horizon of the coupled system, the islands end up behind the horizon of the decoupled system, no matter how we decouple it. This is implied by arguments in \cite{Engelhardt:2014gca}, which use the generalized second law,  that the quantum extremal surface should be behind the horizon.

In the zero temperature case, we needed to consider a very large interval \nref{LargeDis} in order to see the island in the entanglement wedge of a finite interval in the bath region.
If this setup is obtained from approximating a near-extremal black hole in flat space, it is probably necessary to include also a discussion of the soft gravitons \cite{Strominger:2017zoo}.

We also discussed a simple version of the Hawking information paradox involving a two-sided black hole that starts out in the Hartle-Hawking state, and is then evolved forwards in time on both sides.
The geometry is the standard eternal black hole, and there is no modification due to evaporation since the amount of energy emitted in Hawking radiation is the same as falling in from the bath. 
We can consider the entropy of both the baths together.
It starts small, but it linearly increases with time because the black hole evaporates and matter from the bath falls in.
This is the most naive entropy computed for the bath.
Fortunately, the appearance of islands leads to a second quantum extremal surface whose entropy is basically twice the Bekenstein-Hawking entropy. 
This resolves this form of the information paradox.
We get the right answer when applying the right prescription. 
Of course this is the same idea as was described in \cite{Penington:2019npb,Almheiri:2019psf}. 
Our only objective was to present a version of the paradox where the geometry and the quantum state are very simple.

\subsection*{Acknowledgments}
We would like to thank Ying Zhao for initial collaboration and discussions.
We also thank  Raphael Bousso, Hong Zhe Chen, Netta Engelhardt, Tom Faulkner, Daniel Harlow, Tom Hartman, Adam Levine, Henry Maxfield, Rob Myers, Geoffrey Penington, Jorge Santos, Edgar Shaghoulian, Douglas Stanford, Amir Tajdini, Aron Wall, Edward Witten and Zhenbin Yang for useful conversations.
A.A. is supported by funds from the Ministry of Presidential Affairs, UAE.
R.M. is supported by US Department of Energy grant No.\ DE-SC0016244.
J.M. is supported in part by U.S. Department of Energy grant DE-SC0009988 and by the Simons Foundation grant 385600.

\bibliographystyle{apsrev4-1long}
\bibliography{main}
\end{document}